\documentclass[12p,preprint]{revtex4-1}
\usepackage{amsmath,amssymb,graphicx,bm}
\usepackage{epsfig,subfigure}

\begin{document}
\author{Yamen hamdouni}

\title{Quantum mean-field treatment of the dynamics  of a two-level atom in a simple cubic lattice}
\affiliation{Physics department, University Mentouri, Constantine, Algeria}
\email{hamdouniyamen@gmail.com}

\begin{abstract}
The mean field approximation is used to investigate the general features of the dynamics of a two-level atom in a ferromagnetic lattice close to the Curie temperature. Various analytical and numerical results are obtained. We first linearize the lattice Hamiltonian, and we derive the self-consistency equation for the order parameter of the phase transition for arbitrary direction of the magnetic field. The reduced dynamics is deduced by tracing out the degrees of freedom of the lattice, which results in the reduction of the dynamics to that of an atom in an effective spin bath whose size is equal to the size of a unit cell of the lattice. It is found that the dephasing and the excited state occupation probability may be enhanced by applying the magnetic field along some specific directions. The dependence on the change of the temperature  and the magnitude of spin  is also investigated. It turns out that the increase of thermal fluctuations may reduce the occupation probability of the excited state. The entanglement of two such atoms that occupy non-adjacent cells is studied and its variation in time is found to be not much sensitive to the direction of the magnetic field. Entanglement sudden death and revival is shown to occur close to the critical temperature. 
\end{abstract}

\maketitle

\section{Introduction}
The theoretical study of the physical properties of quantum  many-body systems is a challenging task that attracted great amount of attention. These systems  are known to exhibit rather complicated and rich features that lie  at the center of modern applications  such  as those of solid-state  physics \cite{kittel}, spintronics \cite{loss,loss2,loss3}, plasma physics, quantum optics, and quantum information processing  \cite{nielsen, shor, ekert, cirac, mosca, chuang}. Among the most important phenomena that occur in systems with large degrees of freedom, the processes of dissipation, decoherence and dephasing are known to be of greater relevance in either theoretical and practical perspectives. As a matter of fact, the long-standing problem of the transition from the quantum world to the classical world is speculated to be the mere result of the decoherence of the system of interest \cite{zurek1}. Along with that, the dissipation  of energy among the constituents of the composed system plays a crucial role in transport phenomena dealt with in the emerging field of quantum thermodynamics and quantum transport. These facts are in favor of dealing with real quantum systems as open quantum systems \cite{petru}, that are part of or  are interacting with their surrounding in, generally speaking, quite complicated manner. 

The dynamics of a central spin system interacting with its surrounding is a fundamental paradigm that has been investigated by many authors in several contexts \cite{kah,burg,bose,paga,zurek2,ham1,quan,yuan,fischer,ham2,petru2,lai,taka,sarma,lu,zeji,ham3,kici,li}. The most peculiar property of the evolution in spin systems resides in  the non-Markovian nature of the mutual coupling between the central system and its environment; this is manifested in the form of a backflow of information from the environment to the system, due to memory effects, in contrast to the Markovian dynamics \cite{petru}. This is the reason for which a great deal of attention has been given to the problem of characterizing the non-Markovianity in  open quantum systems \cite{rivas,lu2,laine,vasile,fu,hall}.   On the other hand, the problem of the dynamics  of an  impurity in a spin lattice is another example of great relevance. 
In order to facilitate the study of the dynamics in such  lattices, many techniques and approximations have been developed in the course of the last decades, whose applicability and accuracy vary depending on the physical problem at hand. 

It should be stressed that the Green's function technique, which has known a huge success in the theory of many-body systems, represents  one of the most important tools that we dispose \cite{mahan}.  An other method widely used is the   mean-field approximation which has been applied to spin lattices since the early years of the development of the quantum theory of solids, in particular in dealing with phase transitions \cite{majlis}. Recently, the dynamics of a two-level atom impurity that is coupled to a ferromagnetic lattice through Heisenberg XY interaction at zero temperature, has been investigated using spin wave theory and the Green's function technique \cite{ham4}. In this paper, we generalize the investigation to higher temperatures, typically close to the Curie temperature in three dimensions, where spin wave theory ceases to apply. For this purpose, we use the Mean-field approximation \cite{ham5} in order to account for both the thermal fluctuations and their quantum counterpart.

The manuscript is organized as follows. In Section \ref{Sec2}  we introduce the model Hamiltonian for the whole system, and we use the mean-field approximation to linearize the lattice Hamiltonian. There, we derive the self-consistency equation for the order parameter of the phase transition of the lattice, and we also derive the reduced density matrix of the two-level atom. 
  Section \ref{Sec3} is devoted to the study of the dephasing dynamics, where we analyze the evolution in time of the coherences when the atom is prepared in a state that is a linear combination of its ground and excited states, and we investigate the effect of the applied magnetic field and the variation of the temperature on the coherences. Section \ref{Sec4} deals with the populations of the excited and ground states when the atom is initially prepared in the ground state; the dependence of the dynamics on the magnetic field and the the temperature is also analyzed. In Section \ref{sec5} the evolution of the entanglement of two identical two-level atoms, that are located in non-adjacent cells, is studied along with its variation with respect to the model parameters.  We end the paper with a short conclusion.

\section{Model \label{Sec2}}
\subsection{Hamiltonian and mean-field approximation}
We consider a three dimensional   cubic lattice of $N$ spin-$S$ particles, where every particle interacts with its neighboring spins through Heisenberg ferromagnetic interactions. The Hamiltonian of the lattice is explicitly  given  by
\begin{equation}
H_B=-\sum_{\langle i,j \rangle} J_{ij} \bigl(S_i^x S_j^x+S_i^y S_j^y+S_i^z S_j^z\bigr)-\sum_{i=1}^N (h_x S_i^x+h_yS_i^y+h_z S_i^z),
\end{equation}
  where $J_{ij}>0$ denote the  coupling constants, and $\langle i,j \rangle$ indicates that the interactions are restricted to the neighboring spins. In addition, the effect of a constant magnetic field $\vec h$ whose components are designated by $(h_x,h_y,h_z)$ is included. 
	
\begin{figure}[htb]
{\centering{\resizebox*{0.48\textwidth}{!}{\includegraphics{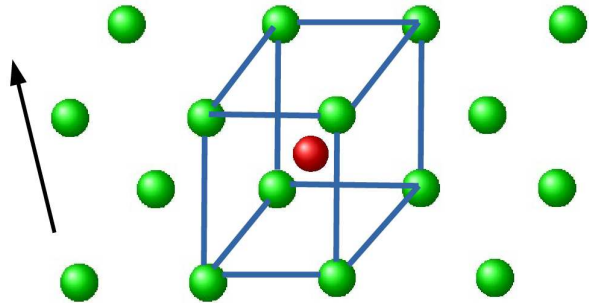}}}
\par}
\caption{Schematic representation of the system studied: a two-level atom located in a unit cell of a ferromagnetic simple cubic lattice interacts with its neighbors. The arrow indicates the direction of the magnetic field.  }\label{fig1}
\end{figure}

	In the center of a unit cell, we place a two-level spin-$\frac{1}{2}$ atom (see Fig.~\ref{fig1}), whose coupling to the lattice  is described by the Hamiltonian
	\begin{equation}
	H_I= S_0^x \sum_{i=1}^\eta\alpha_i S_i^x+S_0^y \sum_{i=1}^\eta\gamma_i S_i^y+S_0^z \sum_{i=1}^\eta\lambda_i S_i^z.
	\end{equation}
	In the above equation,  $S_0^k$ ($k\equiv x,y,z$) are the components of the spin operator of the atom, and  the  constants $\alpha_i$, $\gamma_i$ and $\lambda_i$  are its coupling constants  to the i'th lattice spin. Notice that the summation is meant to be restricted to the $\eta$ spins of the unit cell, namely, $\eta=8$ for the simple cubic lattice. The transition energy between the ground state and the excited state of the atom is denoted by $\omega_0$, such that the free Hamiltonian of the latter reads:
	\begin{equation}
	 H_0=\omega_0 S_0^z.
	\end{equation}
	
	 We assume that the lattice is in thermal equilibrium at temperature $T$. In the absence of the magnetic field,   the ferromagnetic interactions tend to point the spins in the same direction up to the Curie temperature of the lattice, above which the latter undergoes a phase transition to the paramagnetic phase.  When an external magnetic field is applied, the ferromagnetic phase  is stabilized at even higher temperatures. In general, the phase transition is well described by its order parameter $m$. For the present model, assuming translation invariance,  the order parameter is given by
	\begin{equation}
	m=\sqrt{m_x^2+m_y^2+m_z^2}
	\end{equation}
	where
	\begin{equation}
	 \vec m= \Bigl\langle \vec S_i \Bigr\rangle.
	\end{equation}
         Here, the expectation value is evaluated with respect to the Gibbs equilibrium state of the lattice, that is:
	\begin{equation}
	\Bigl\langle \vec S_i \Bigr\rangle =\mathrm{tr} \bigl(\vec S_i \  e^{-\beta H_B}/Z\Bigr),
	\end{equation}
	with $\beta$ being the inverse temperature, and $Z$ is  the partition function of the lattice:
	\begin{equation}
	Z =\mathrm{tr} \bigl(e^{-\beta H_B}).
	\end{equation}
	
	We now invoke the mean-field approximation  which  enables us to linearize the lattice Hamiltonian. The essence of latter  approximation consists in neglecting higher order deviations from the mean values of the spin operators. As a result,  the mean-field Hamiltonian of the lattice reads:
	\begin{eqnarray}
	H_B^{\rm mf}&=&\sum_{\langle i,j \rangle} J_{ij} m^2-2\sum_{\langle i,j \rangle} J_{ij}\Bigl(m_x S_i^x+m_y S^y_i+m_z S_i^z\Bigr)\nonumber \\ &-&\sum_{i=1}^N (h_x S_i^x+h_yS_i^y+h_z S_i^z).
	\end{eqnarray}
	Since we assumed translation invariance, we may further write:
	\begin{eqnarray}
	H_B^{\rm mf}&=&N \sum_{\vec \delta} J_{\vec\delta} m^2-2\sum_{i=1}^N \sum_{\vec \delta} J_{\vec\delta}\Bigl(m_x S_i^x+m_y S^y_i+m_z S_i^z\Bigr)\nonumber \\ &-&\sum_{i=1}^N (h_x S_i^x+h_yS_i^y+h_z S_i^z),
	\end{eqnarray}
	where $\sum_{\vec \delta}$ is the sum with respect to all vectors $\vec\delta$ linking a given lattice spin to its neighbors. The latter Hamiltonian can be put into the compact form
	\begin{equation}
	H_B^{\rm mf}=N \sum_{\vec \delta} J_{\vec\delta} m^2-\Biggl(2\vec m \sum_{\vec \delta} J_{\vec\delta}+\vec h\Biggr) \sum_{i=1}^N \vec S_i.
	\end{equation}
	
	The partition function corresponding to the above Hamiltonian reads:
	\begin{equation}
	Z=e^{-\beta N \sum_{\vec \delta} J_{\vec\delta} m^2}\prod_{i=1}^N \mathrm {tr} \ e^{\beta \Biggl(2\vec m \sum_{\vec \delta} J_{\vec\delta}+\vec h\Biggr) \vec S_i}.
	\end{equation}
	By a unitary transformation, we can show that the partition function is calculated as:
	
\begin{eqnarray}
&&Z= \exp\{-\beta N \sum_{\vec \delta} J_{\vec\delta} m^2\}\nonumber \\ & &\times\Biggl[1+\frac{2\cosh \Bigl[\frac{\beta (S+1)}{2}\|2\vec m \sum_{\vec \delta} J_{\vec\delta}+\vec h\|\Bigr] \sinh \Bigl[\frac{\beta S}{2}\|2\vec m \sum_{\vec \delta} J_{\vec\delta}+\vec h\|\Bigr]}{\sinh \Bigl[\frac{\beta}{2}\|2\vec m \sum_{\vec \delta} J_{\vec\delta}+\vec h\|\Bigr]}\Biggr]^N.
\end{eqnarray}
	The free energy per spin  of the lattice is given in terms of the partition function by
	\begin{equation}
	 F=-1/(N\beta) \ln Z.
	\end{equation}
	By minimizing the free energy with repect to $m_x$, $m_y$ and $m_z$, we obtain the following self-consistency equations:
\begin{eqnarray}
m_x&=& \frac{h_x+2 m_x \sum_{\vec \delta } J_{\vec\delta}}{\Lambda}\Biggl\{ \cosh\Bigl[\frac{\beta}{2} (S+1)\Lambda\Bigr]\Biggl[S\cosh(\beta \Lambda S)-\coth(\frac{\beta}{2} \Lambda) \sinh(\frac{\beta}{2} S\Lambda)\Biggr]\nonumber \\
&+&(1+S) \sinh(\frac{\beta}{2} S\Lambda) \sinh(\frac{\beta}{2} (S+1)\Lambda)\Biggr\}\frac{1}{\sinh(\frac{\beta}{2}(1+2 S)\Lambda)},\\
m_y&=& \frac{h_y+2 m_y \sum_{\vec \delta } J_{\vec\delta}}{\Lambda}\Biggl\{ \cosh\Bigl[\frac{\beta}{2} (S+1)\Lambda\Bigr]\Biggl[S\cosh(\beta \Lambda S)-\coth(\frac{\beta}{2} \Lambda) \sinh(\frac{\beta}{2} S\Lambda)\Biggr]\nonumber \\
&+&(1+S) \sinh(\frac{\beta}{2} S\Lambda) \sinh(\frac{\beta}{2} (S+1)\Lambda)\Biggr\}\frac{1}{\sinh(\frac{\beta}{2}(1+2 S)\Lambda)},
\\
m_z&=& \frac{h_z+2 m_z \sum_{\vec \delta } J_{\vec\delta}}{\Lambda}\Biggl\{ \cosh\Bigl[\frac{\beta}{2} (S+1)\Lambda\Bigr]\Biggl[S\cosh(\beta \Lambda S)-\coth(\frac{\beta}{2} \Lambda) \sinh(\frac{\beta}{2} S\Lambda)\Biggr]\nonumber \\
&+&(1+S) \sinh(\frac{\beta}{2} S\Lambda) \sinh(\frac{\beta}{2} (S+1)\Lambda)\Biggr\}\frac{1}{\sinh(\frac{\beta}{2}(1+2 S)\Lambda)}.
\end{eqnarray}
where for ease of notation we have  introduced the quantity $$\Lambda=\|2\vec m \sum_{\vec \delta} J_{\vec\delta}+\vec h\|.$$
The calculation of the components $m_j$ is sufficient to fix the form of the mean-field Hamiltonian. From the above equations we infer that the order parameter $m$ may be determined self-consistently through the equation: 
\begin{eqnarray}
m&=&\Biggl\{ \cosh\Bigl[\frac{\beta}{2} (S+1)\Lambda\Bigr]\Biggl[S\cosh(\beta \Lambda S)-\coth(\frac{\beta}{2} \Lambda) \sinh(\frac{\beta}{2} S\Lambda)\Biggr]\nonumber \\
&+&(1+S) \sinh(\frac{\beta}{2} S\Lambda) \sinh(\frac{\beta}{2} (S+1)\Lambda)\Biggr\}\frac{1}{\sinh(\frac{\beta}{2}(1+2 S)\Lambda)}\label{self}.
\end{eqnarray}

	Equation (\ref{self}) enables us to derive the Curie temperature of the lattice in zero external field, namely:
	\begin{equation}
	T_c=\frac{2S(S+1) \sum_{\vec \delta} J_{\vec\delta}}{3k_B}.
	\end{equation}
	As was pointed out earlier, in zero magnetic field the lattice exhibits ferromagnetic order bellow the Curie temperature; however, when the magnetic field is applied,  the  ferromagnetic phase persists at temperatures larger than $T_c$. The order of these temperatures depends on the strength of the magnetic field, and cannot be established directly from  Eq. (\ref{self}), since for nonzero $\vec h$, the latter equation has always a non-zero solution for the order parameter $m$. That is the reason for which we shall restrict ourselves from here on to temperatures lower than the Curie temperature $T_c$.

	\subsection{Reduced dynamics}
	
	The evolution of the density matrix describing the state of the atom is given by
	\begin{equation}
	\rho(t)=\frac{1}{Z}\mathrm{tr_B} e^{-it (H_0+H_B+H_I)} \ \rho(0)\otimes e^{-\beta H_B} e^{it (H_0+H_B+H_I)},
	\end{equation}
	where $\mathrm{tr_B}$ designates the trace with respect to the lattice degrees of freedom, and $\rho(0)$ is the initial state of the atom which is assumed initially uncorrelated with that of the lattice. Taking into account the mean-field Hamiltonian, we may write:
		\begin{eqnarray}
		\rho(t)&=&\frac{1}{Z} \mathrm{tr_B} e^{-it \bigl(\omega_0 S_0^z+ \sum\limits_{i=1}^N \vec\Gamma_i \vec S_i\bigr)}  \ \rho(0)\otimes e^{\beta \Bigl(2m \frac{\sum_{\vec \delta} J_{\vec\delta}}{\|\vec h\|}+1\Bigr) \sum\limits_{i=1}^N \vec h\vec S_i} e^{it \bigl(\omega_0 S_0^z+ \sum\limits_{i=1}^N \vec\Gamma_i \vec S_i\bigr)} 
		\label{den1}
		\end{eqnarray}
		where we use the notations:
		\begin{eqnarray}
		\vec\Gamma_i &=&-\Bigl(2m \frac{\sum_{\vec \delta} J_{\vec\delta}}{\|\vec h\|}+1\Bigr)\vec h+\vec K_i,\\
		\vec K_i&=&(\alpha_i S_0^x,\gamma_i S_0^y,\lambda_i S_0^z).
		\end{eqnarray}
		When the coupling of the atom is restricted to its nearest neighbors, we should have:
	\begin{equation}
 \vec K_i=\Biggl\{ \begin{array} {c}
\vec K= (\alpha S_0^x,\gamma S_0^y,\lambda S_0^z) \quad {\rm for} \quad i\leq\eta, \\
  \vec  0 \qquad \qquad\qquad {\rm for} \quad i>\eta,
   \end{array}
\end{equation}
where $\alpha$, $\gamma$ and $\lambda$ are the nearest-neighbor coupling constants. This implies that
		
		\begin{equation}
 \vec \Gamma_i=\Biggl\{ \begin{array} {c}
\vec \Gamma= -\Bigl(2m \frac{\sum_{\vec \delta} J_{\vec\delta}}{\|\vec h\|}+1\Bigr)\vec h+\vec K \quad {\rm for} \quad i\leq\eta, \\
  -\Bigl(2m \frac{\sum_{\vec \delta} J_{\vec\delta}}{\|\vec h\|}+1\Bigr)\vec h \qquad\qquad \quad{\rm for} \quad i>\eta.
   \end{array}
\end{equation}
Afterward, by 	taking the trace in Eq.(\ref{den1}), we end up with:
	\begin{eqnarray}
	\rho(t)&=&\frac{1}{\widetilde Z} \mathrm{tr_{\widetilde B}} e^{-it \bigl(\omega_0 S_0^z+ \vec\Gamma \sum\limits_{i=1}^\eta  \vec S_i\bigr)}  \ \rho(0)\otimes e^{\beta \Bigl(2m \frac{\sum_{\vec \delta} J_{\vec\delta}}{\|\vec h\|}+1\Bigr) \sum\limits_{i=1}^\eta \vec h\vec S_i} e^{it \bigl(\omega_0 S_0^z+ \vec\Gamma \sum\limits_{i=1}^\eta  \vec S_i\bigr)}\label{den2}
		\end{eqnarray}
	with	
\begin{eqnarray}
&&\widetilde Z= \exp\{-\beta \eta \sum_{\vec \delta} J_{\vec\delta} m^2\}\nonumber \\ & &\times\Biggl[1+\frac{2\cosh \Bigl[\frac{\beta (S+1)}{2}\Bigl(2m \sum_{\vec \delta} J_{\vec\delta}+\|\vec h\|\Bigr)\Bigr] \sinh \Bigl[\frac{\beta S}{2}\Bigl(2m \sum_{\vec \delta} J_{\vec\delta}+\|\vec h\|\Bigr)\Bigr]}{\sinh \Bigl[\frac{\beta}{2}\Bigl(2m \sum_{\vec \delta} J_{\vec\delta}+\|\vec h\|\Bigr)\Bigr]}\Biggr]^\eta,
\end{eqnarray}
whereas  $	\mathrm{tr}_{\widetilde  B} $ denotes the trace with respect to the spins in the unit cell where the atom is located.
	
Consequently, the atom behaves as if it is interacting with an effective spin bath whose size is precisely the size of the aforementioned  cell.  The overall effect of the total lattice on the atom is encoded in the order parameter $m$, along with the coupling constants $J_\delta$,  and the Curie temperature $T_c$. In the next section, we investigate the variation of the reduced density matrix and its dependence on the relevant parameters of the model, in particular the magnetic field and the temperature.

	\section{Pure dephasing \label{Sec3}}
	In the particular case when $\alpha=\gamma=0$, which corresponds to the Ising interaction Hamiltonian
	\begin{equation}
	 H_I=\lambda S_0^z \sum\limits_{i=1}^\eta S_i^z,
	\end{equation}
	the central atom does not exchange energy with the lattice. The diagonal elements of its density matrix do not evolve in the course of the time, but preserve their initial values; this implies that  for the present parameters of the model, both the ground and the excited states $|g\rangle$ and $|e\rangle$  of the atom are not affected by the lattice.  The coherences in general do change with time, leading to pure dephasing dynamics of the atom. We shall mainly be interested in the dynamics in a simple cubic lattice, with nearest-neighbor coupling for which $\eta=8$, and the Curie temperature reads
	\begin{equation}
	T_c=\frac{4S(S+1)  J}{k_B}
	\end{equation}
	where $J$ refers to the nearest-neighbor coupling constant.
	\begin{figure}[htb]
{\centering\subfigure[{}]{\resizebox*{0.33\textwidth}{!}{\includegraphics{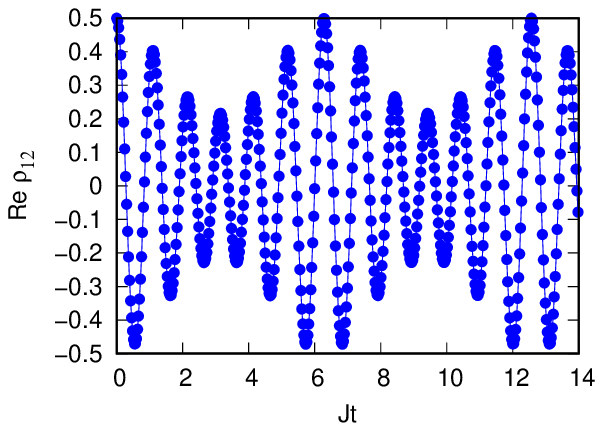}}}
\subfigure[{}]{
\resizebox*{0.33\textwidth}{!}{\includegraphics{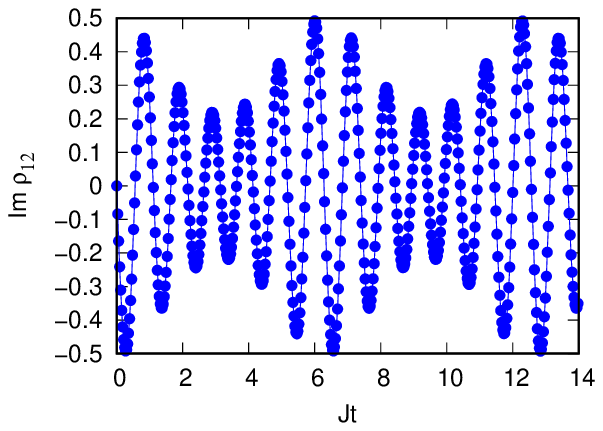}}}
\subfigure[{}]{
\resizebox*{0.33\textwidth}{!}{\includegraphics{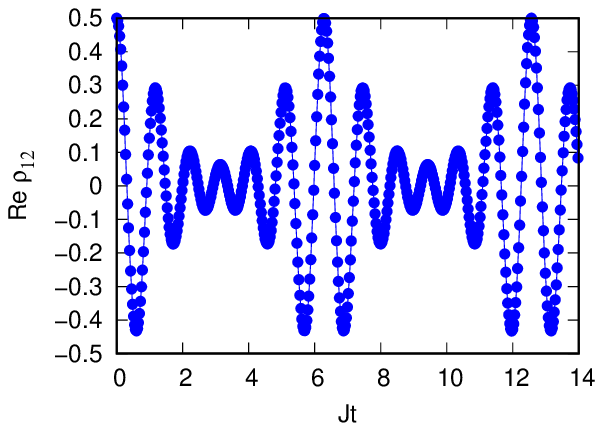}}}
\subfigure[{}]{
\resizebox*{0.33\textwidth}{!}{\includegraphics{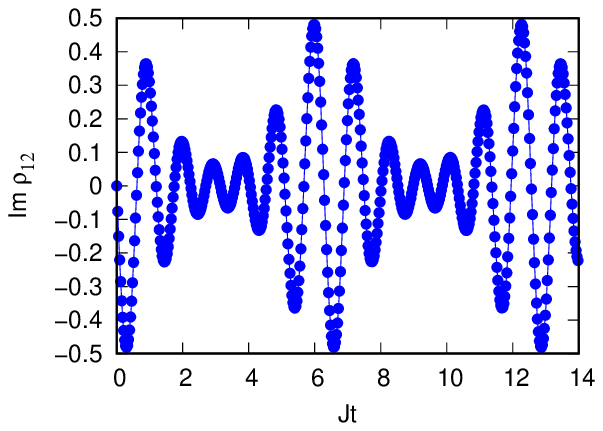}}}
\par}
\caption{The evolution of the elements of the reduced density matrix corresponding to the initial state $\phi(0)$. For (a) and (b): $h_z$=0.5 $J$, $h_x=h_y=0$, $T=2J$. For (c)  and (d):  $h_z$=0.5 $J$, $h_x=h_y=0$, $T=2.5J$  . The other parameters are  $S=1/2$, $\omega_0=2J$, $\alpha=\gamma=0$, $\lambda=J$.}\label{fig2}
\end{figure}
	Let us begin with a  two-level  atom that is embedded in a lattice of spin-$\frac{1}{2}$ atoms, that is $S=S_0=\frac{1}{2}$. It can be shown that the off-diagonal element of the reduced density matrix evolves according to:
	
\begin{eqnarray}
\rho_{12}(t)=\rho_{12}(0) e^{-i \omega_0 t} F(t)^\eta
\end{eqnarray}
	with
	\begin{eqnarray}
&&F(t)=\frac{1}{2\cosh\bigl(\frac{\beta}{2}(12 J m +\|\vec h\|)\bigr)}\mathrm{tr}\Biggl[\biggl (\cos(\frac{t\Lambda_-}{2})+\frac{2i}{\Lambda_-} \sin(\frac{t\Lambda_-}{2}) H_-\biggr)\nonumber\\ &&\times\biggl(\cosh(\beta\Lambda/2)+\frac{2}{\Lambda}\sinh(\beta\Lambda/2)  H_b\biggr)\biggl(\cos(\frac{t\Lambda_+}{2})-\frac{2i}{\Lambda_+} \sin(\frac{t\Lambda_+}{2}) H_+\biggr)\ \Biggr]
\end{eqnarray}
where the operators $H_b$ and $H_\pm$ along with the quantities $\Lambda_\pm$ are given by:
\begin{eqnarray}
H_b&=&\Bigl(\frac{12 J m }{\|\vec h\|}+1\Bigr) \vec h\vec S,\\
H_\pm&=&\Bigl(\frac{12 J m }{\|\vec h\|}+1\Bigr) \vec h\vec S\pm \frac{\lambda}{2} S_z,\\ 
\Lambda_\pm&=&\sqrt{(12m J + \|\vec h\|)^2+\frac{\lambda^2}{4}\pm\lambda\Bigl(\frac{12 J m }{\|\vec h\|}+1\Bigr)}.
\end{eqnarray}

\begin{figure}[htb]
{\centering\subfigure[{}]{\resizebox*{0.33\textwidth}{!}{\includegraphics{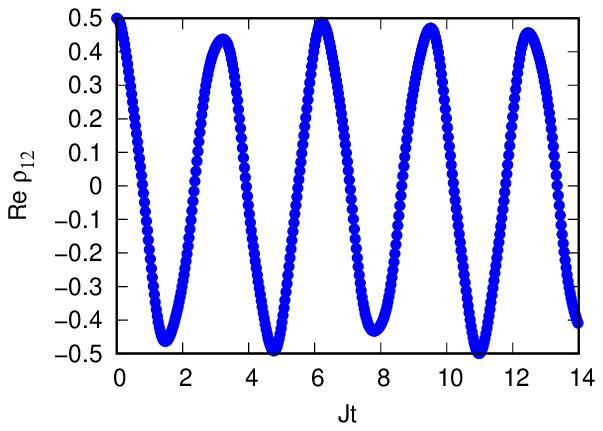}}}
\subfigure[{}]{
\resizebox*{0.33\textwidth}{!}{\includegraphics{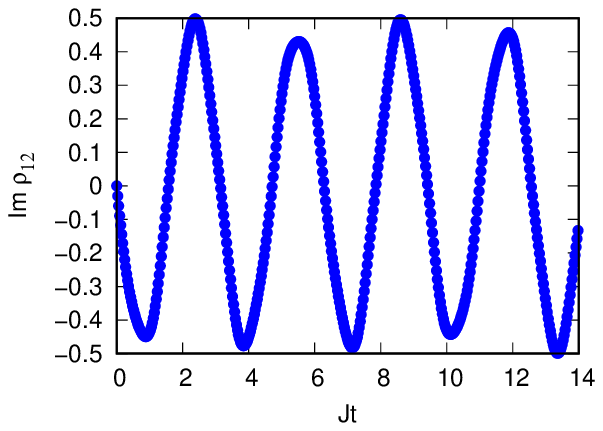}}}
\subfigure[{}]{
\resizebox*{0.33\textwidth}{!}{\includegraphics{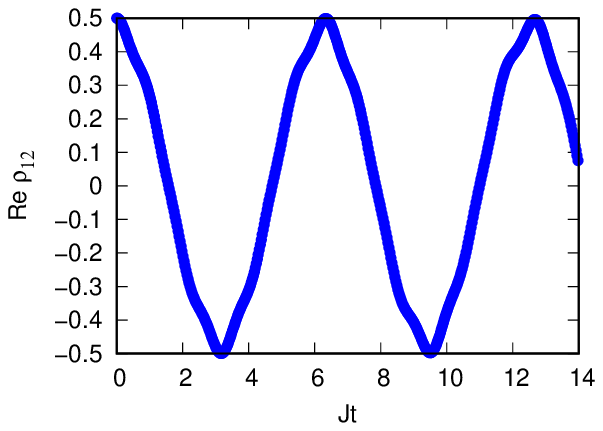}}}
\subfigure[{}]{
\resizebox*{0.33\textwidth}{!}{\includegraphics{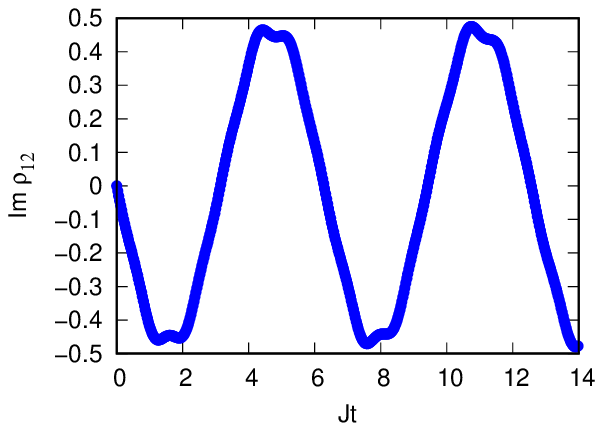}}}
\par}
\caption{The evolution of the elements of the reduced density matrix corresponding to the initial state $\phi(0)$. For (a) and (b): $h_x$=0.5 $J$, $h_z=h_y=0$, $\omega_0=2J$. For (c)  and (d):  $h_x$=0.5 $J$, $h_z=h_y=0$, $\omega_0=J$. The other parameters are  $S=1/2$, $T=2J$, $\alpha=\gamma=0$, $\lambda=J$.}\label{fig3}
\end{figure}

\begin{figure}[htb]
{\centering\subfigure[{}]{\resizebox*{0.33\textwidth}{!}{\includegraphics{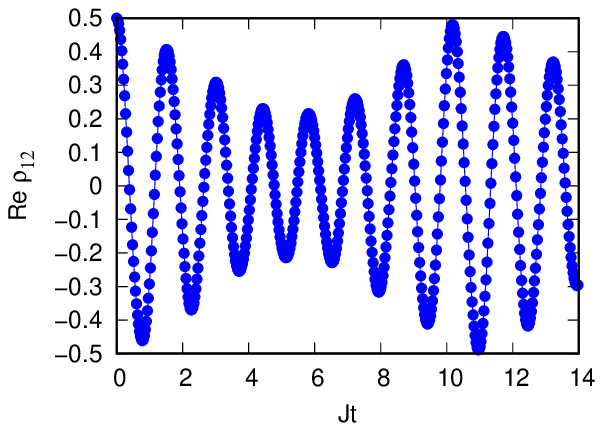}}}
\subfigure[{}]{
\resizebox*{0.33\textwidth}{!}{\includegraphics{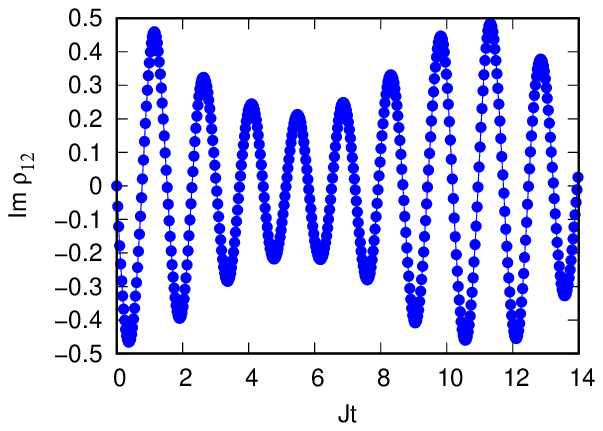}}}
\subfigure[{}]{
\resizebox*{0.33\textwidth}{!}{\includegraphics{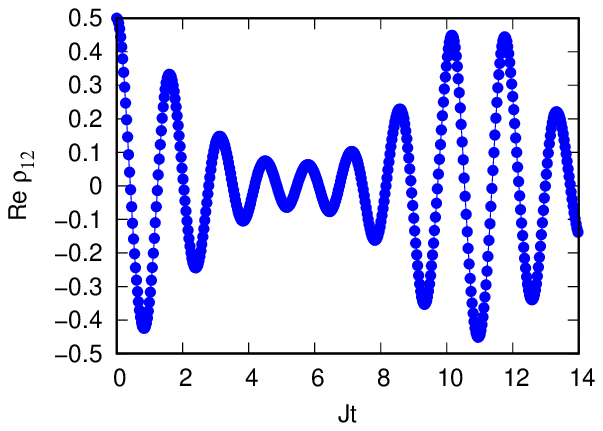}}}
\subfigure[{}]{
\resizebox*{0.33\textwidth}{!}{\includegraphics{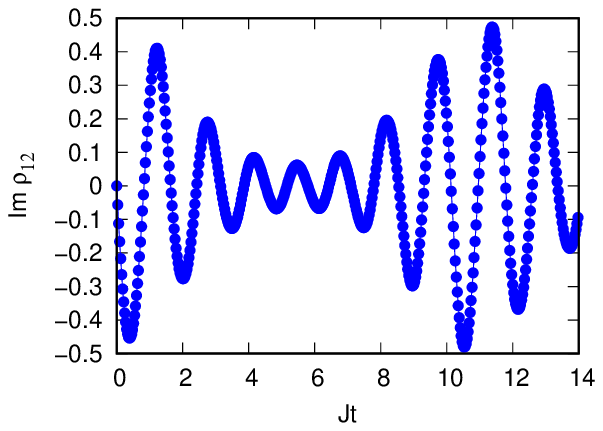}}}
\par}
\caption{The  variation of the elements of the reduced density matrix  corresponding to the initial state $\phi(0)$. For (a) and (b): $h_x=h_y=h_z=J/2\sqrt{3}$, $T=2J$. For (c)  and (d): $h_x=h_y=h_z=J/2\sqrt{3}$, $T=2.5J$. The other parameters are  $S=1/2$, $\omega_0=2J$, $\alpha=\gamma=0$, $\lambda=J$.}\label{fig4}
\end{figure}

\begin{figure}[htb]
{\centering\subfigure[{}]{\resizebox*{0.33\textwidth}{!}{\includegraphics{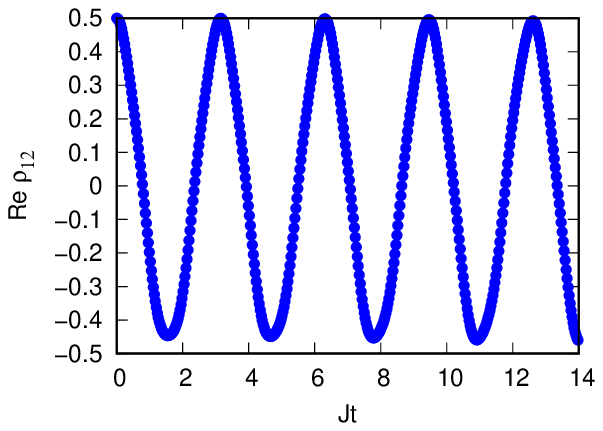}}}
\subfigure[{}]{
\resizebox*{0.33\textwidth}{!}{\includegraphics{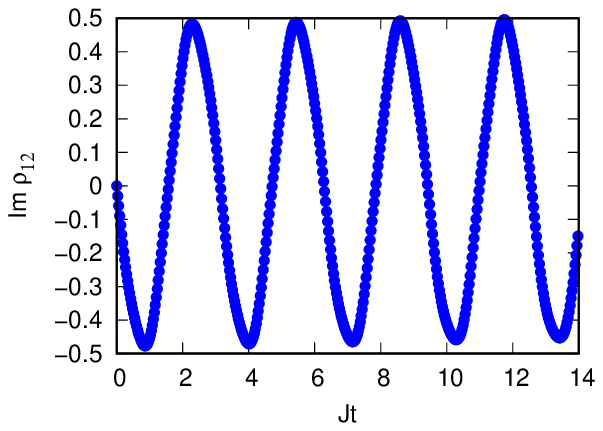}}}
\par}
\caption{The  variation of the elements of the reduced density matrix  corresponding to the initial state $\phi(0)$ for $h_x=h_y=J/2\sqrt{2}$, $T=2J$,  $S=1/2$, $\omega_0=2J$, $h_z=0$,  $\alpha=\gamma=0$, $\lambda=J$.}\label{fig5}
\end{figure}

 In Fig. \ref{fig2}  we display the real and the imaginary parts of $\rho_{12}$
when the atom is initially prepared in the state
 \begin{equation}
  \phi(0)=\frac{1}{\sqrt{2}}(|g\rangle+|e\rangle)
 \end{equation}
 The  plots correspond to a lattice with $S=\frac{1}{2}$  for uniform  coupling of the atom at different temperatures. The magnetic field in this case is chosen in the $z$ direction. 
 It can be seen that the coherence exhibits oscillatory decay and revival, which indicates the significance of the lattice memory effects on the dynamics of the atom that persist even  at high temperature where the  thermal fluctuations are much important.  We recall that the state of a two-level atom  may be described by the density matrix:
\begin{equation}
\rho(t)=\frac{1}{2}\begin{pmatrix}1+r_3(t)& r_1(t)-ir_2(t)\\
r_1(t)+ir_2(t)&1-r_3(t)\end{pmatrix},\end{equation}
where  $(r_1,r_2,r_3)$ are the components of the Bloch vector 	 $\vec{r}$. The latter  satisfies the condition $|\vec{r}|\leq 1$, with the equality being verified if the  state is pure. 
Within this representation, the expectation values of the components of the spin operator of the atom at any moment $t$ are $\langle S_0^x \rangle(t)= \frac{ r_1(t)}{2}$,  $\langle S_0^y \rangle(t)= \frac{ r_2(t)}{2}$, and $\langle S_0^z \rangle(t)= \frac{ r_3(t)}{2}$.
Hence, the initial state $\phi(0)$ evolves in the subsequent times to a state for which
\begin{eqnarray}
	\langle S_0^x \rangle(t)&=& \frac{1 }{2}[ \cos(\omega_0 t) {\rm Re}\  F(t)^\eta+ \sin(\omega_0 t) {\rm Im}\  F(t)^\eta],\\
	\langle S_0^y \rangle(t)&=& \frac{1 }{2}[ \sin(\omega_0 t) {\rm Re}\  F(t)^\eta- \cos(\omega_0 t) {\rm Im}\  F(t)^\eta],\\
	\langle S_0^z \rangle(t)&=& 0.
	\end{eqnarray}
	 We notice that the evolved state approaches in multiple instances the pure initial state. However, at high temperature the lapses of the time in which the atom evolves close to the fully mixed state are much larger. An other point worth observing is that the state of the atom  evolves periodically close  to the pure states $\psi=\frac{1}{\sqrt{2}}(|{g}\rangle\pm i|e\rangle)$. It turns out that for stronger values of the magnetic field, the periods of the oscillations decrease, and the decay and revival is much faster for both low and high temperatures.
	
	A quite different  situation occurs when the magnetic field points in the $x$ direction, see figure \ref{fig3}.  Indeed, in this case the frequencies of oscillations have significantly decreased, and  it turns out that the dynamics is much less sensitive to the variation of the temperature. Increasing the latter does not lead to a noticeable decrease of the off-diagonal elements. In order to check the significance of thermal fluctuations, we also plot the variation of the matrix elements for a smaller value of the proper energy $\omega_0$, leaving the other parameters unchanged. We find  that the bar effect resides in a decrease in the frequency of the collapse and  revival of the coherence; again, the dynamics is not affected much by the variation of the temperature.     When the magnetic field points along the direction (1,1,1), we notice that the decoherence is again more sensitive to the temperature, but the revival of the coherence takes larger time to occur as displayed in figure \ref{fig4}. This leads to the conclusion that as long as the component of the magnetic field along the $z$-direction is nonzero, the dynamics displays a strong dependence on the temperature. To verify this assertion, we show  in figure \ref{fig5} the case when the magnetic field points within the $x-y$ plane. It is clear that this situation  is quite similar to that corresponding to  a magnetic field pointing in the $x$ direction.
	
	\begin{figure}[htb]
{\centering\subfigure[{}]{\resizebox*{0.33\textwidth}{!}{\includegraphics{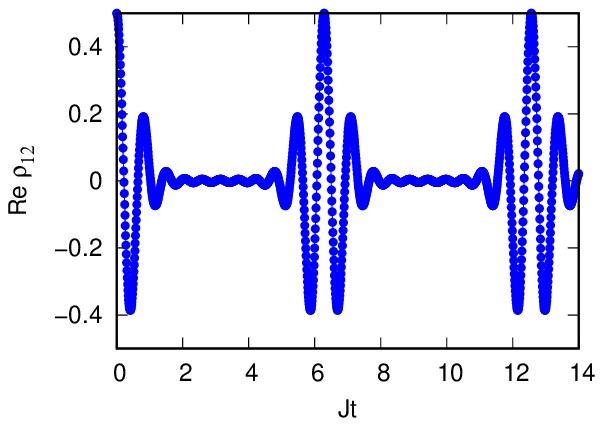}}}
\subfigure[{}]{
\resizebox*{0.33\textwidth}{!}{\includegraphics{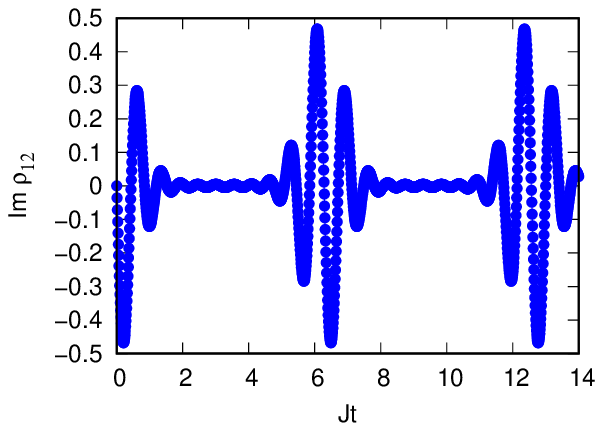}}}
\subfigure[{}]{
\resizebox*{0.33\textwidth}{!}{\includegraphics{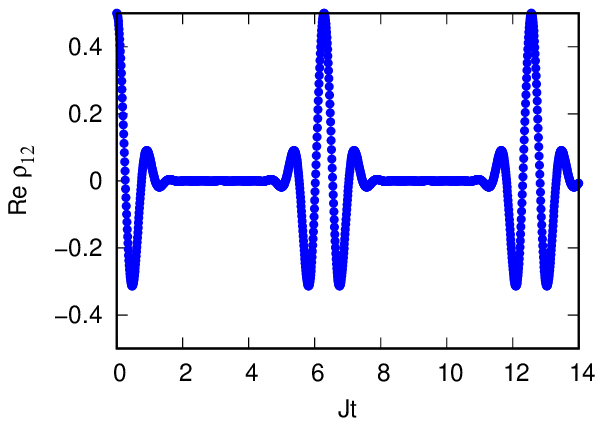}}}
\subfigure[{}]{
\resizebox*{0.33\textwidth}{!}{\includegraphics{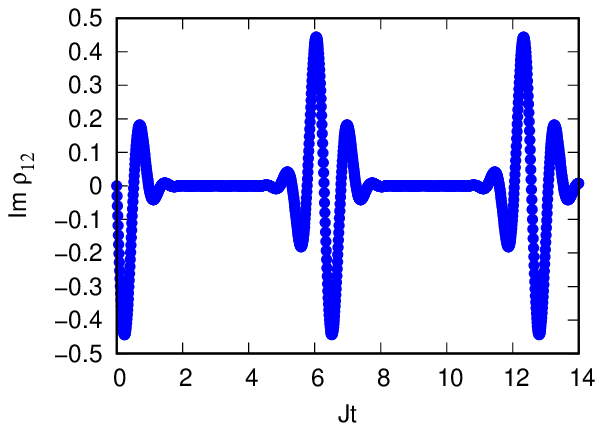}}}
\par}
\caption{The evolution of the elements of the reduced density matrix corresponding to the initial state $\phi(0)$. For (a) and (b): $h_z$=0.5 $J$, $h_x=h_y=0$, $T=7J$. For (c)  and (d):  $h_z$=0.5 $J$, $h_x=h_y=0$, $T=7.8J$  . The other parameters are  $S=1$, $\omega_0=2J$, $\alpha=\gamma=0$, $\lambda=J$}\label{fig6}
\end{figure}	
	
	For a two-level atom in a lattice whose atoms have spin $S=1$, we find that
	\begin{eqnarray}
F(t)&=&\frac{1}{1+2\cosh\bigl(\beta(12m J +|\vec h|)\bigr)}\mathrm{tr}\Biggl[\biggl(1+\frac{1}{\Lambda_-}(\cos(t\Lambda_-)-1)H_-\nonumber\\&+&\frac{i}{\Lambda_-^2} \sin(t\Lambda_-) H_-^2\biggr)\biggl(1+\frac{1}{\Lambda}(\cosh(\beta\Lambda)-1)H_b+\frac{1}{\Lambda^2} \sinh(\beta\Lambda) H_b^2\biggr)\nonumber \\&\times&\biggl(1+\frac{1}{\Lambda_+}(\cos(t\Lambda_+)-1)H_+-\frac{i}{\Lambda_+^2} \sin(t\Lambda_+) H_+^2\biggr)\ \Biggr]
\end{eqnarray}
	\begin{figure}[tbh]
{\centering\subfigure[{}]{\resizebox*{0.33\textwidth}{!}{\includegraphics{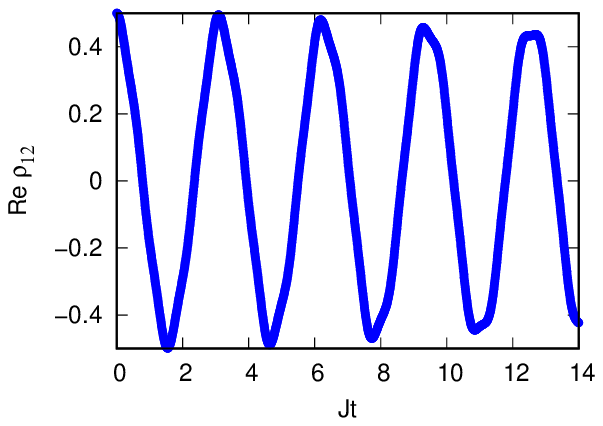}}}
\subfigure[{}]{
\resizebox*{0.33\textwidth}{!}{\includegraphics{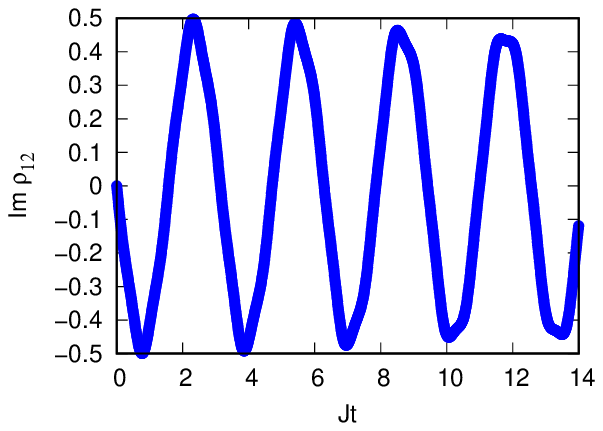}}}
\subfigure[{}]{
\resizebox*{0.33\textwidth}{!}{\includegraphics{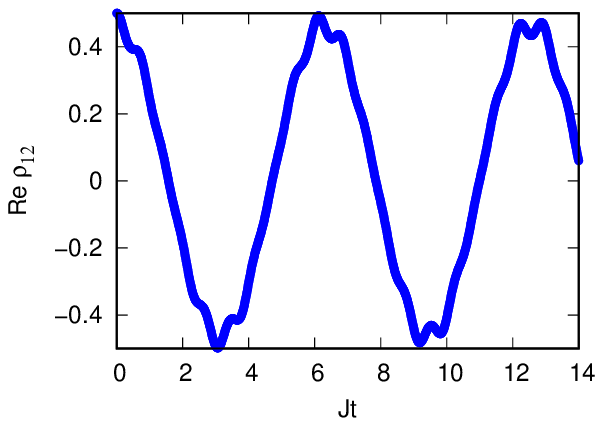}}}
\subfigure[{}]{
\resizebox*{0.33\textwidth}{!}{\includegraphics{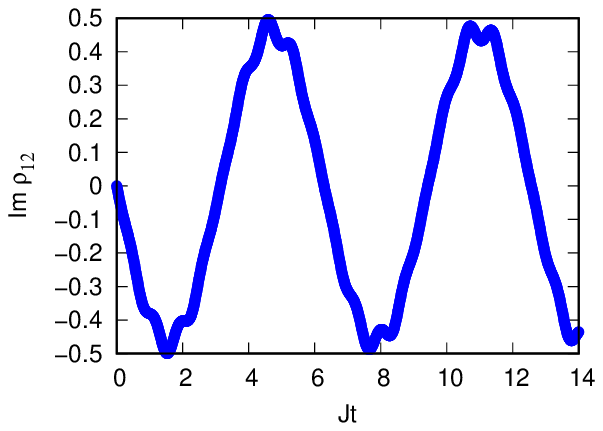}}}
\par}
\caption{The evolution of the elements of the reduced density matrix corresponding to the initial state $\phi(0)$. For (a) and (b): $h_x$=0.5 $J$, $h_z=h_y=0$, $\omega_0=2J$. For (c)  and (d):  $h_x$=0.5 $J$, $h_z=h_y=0$, $\omega_0=J$. The other parameters are  $S=1$, $T=7J$, $\alpha=\gamma=0$, $\lambda=J$.}\label{fig7}
\end{figure}

\begin{figure}[tbh]
{\centering\subfigure[{}]{\resizebox*{0.33\textwidth}{!}{\includegraphics{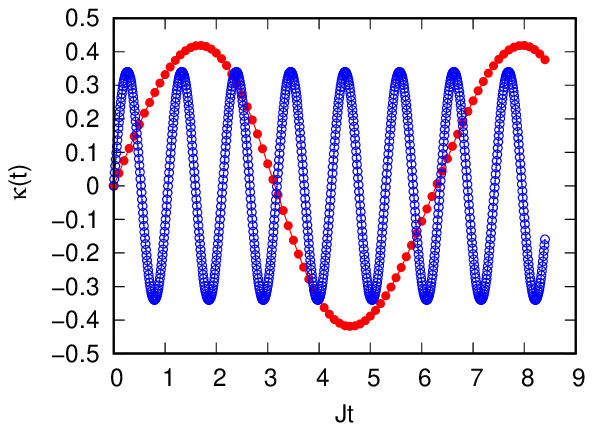}}}
\subfigure[{}]{
\resizebox*{0.33\textwidth}{!}{\includegraphics{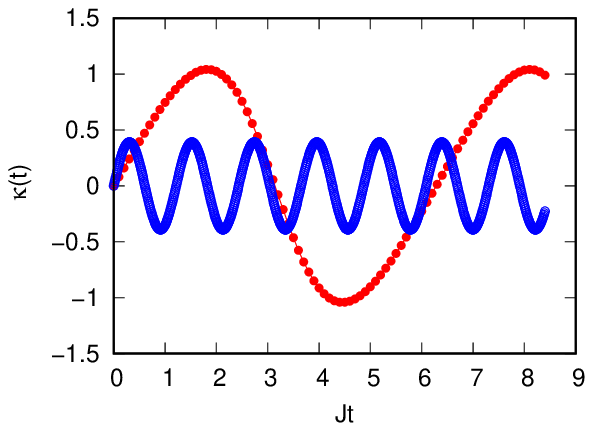}}}
\subfigure[{}]{
\resizebox*{0.33\textwidth}{!}{\includegraphics{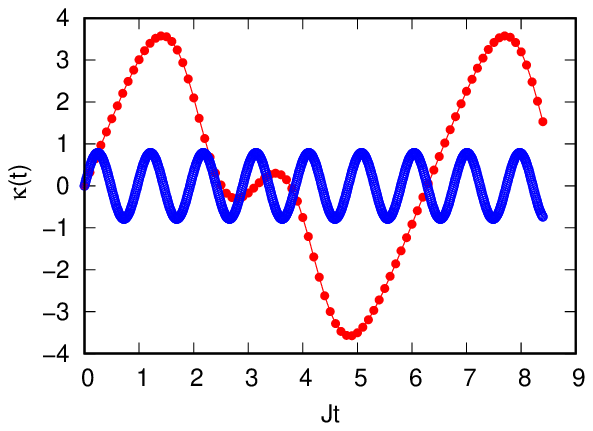}}}
\subfigure[{}]{
\resizebox*{0.33\textwidth}{!}{\includegraphics{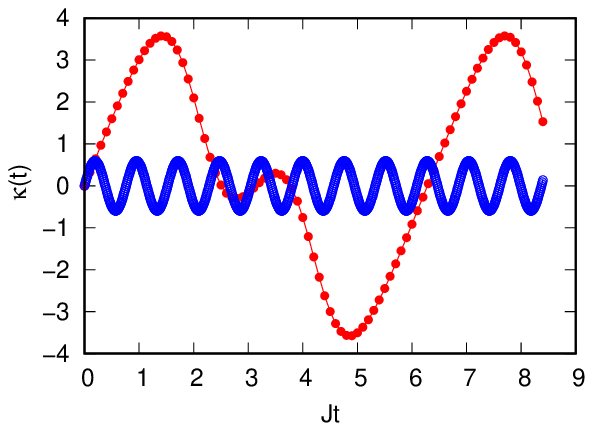}}}
\par}
\caption{The dephasing rate as a function of time for different values of the temperature and the magnetic field. For (a): $S=\frac{1}{2}$, $h_z=0.5 J$ (slowly oscillating curve), $h_x=0.5$ (quickly oscillating curve), and  $T=2J$. For (b):  $S=\frac{1}{2}$, $h_z=0.5 J$ (slowly oscillating curve), $h_x=0.5$ (quickly oscillating curve) and  $T=2.5J$. For (c):  $S=1$, $h_z=0.5 J$ (slowly oscillating curve), $h_x=0.5$ (quickly oscillating curve) and  $T=7J$. For (d):  $S=1$, $h_z=0.5 J$ (slowly oscillating curve), $h_x=0.5$ (quickly oscillating curve) and $T=7.8J$. The other parameters are $h_y=0$, $\alpha=\gamma=0$ and $\lambda=J$.}\label{fig8}
\end{figure}
	Figures \ref{fig6} and \ref{fig7} display the evolution in time of the coherences when the lattice is composed of spin-1 particles for the  directions of the magnetic fields corresponding to figures \ref{fig2} and \ref{fig4} respectively, for  comparison with the case of spin-$\frac{1}{2}$. We clearly notice that as the temperature gets closer to the Curie temperature of the lattice, the loss of coherence is much important  when the magnetic field is applied along the $z$-direction; actually for temperatures high enough the state of the atom becomes completely mixed during relatively long intervals of time. Despite this, there occurs periodic revival of the coherence, which demonstrates,  once again, the significance of  the purely quantum memory effects of the lattice which are responsible of the backflow of information from the lattice to the atom even when the thermal fluctuations dominate. However, when the magnetic field points along the $x$-direction, the decay of the coherences is much very slow and they take on relatively large values periodically in the course of the time.
	\begin{figure}[tbh]
{\centering\subfigure[{}]{\resizebox*{0.33\textwidth}{!}{\includegraphics{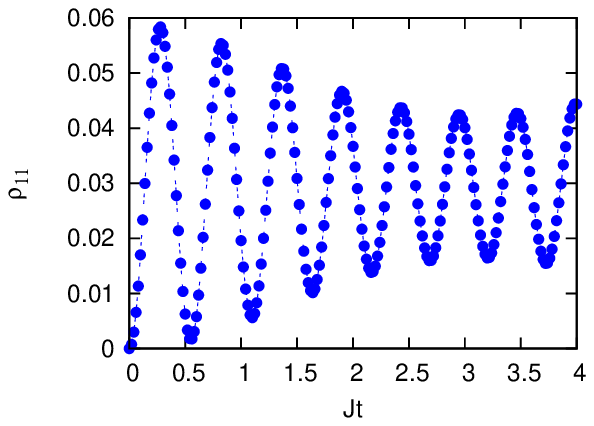}}}
\subfigure[{}]{
\resizebox*{0.33\textwidth}{!}{\includegraphics{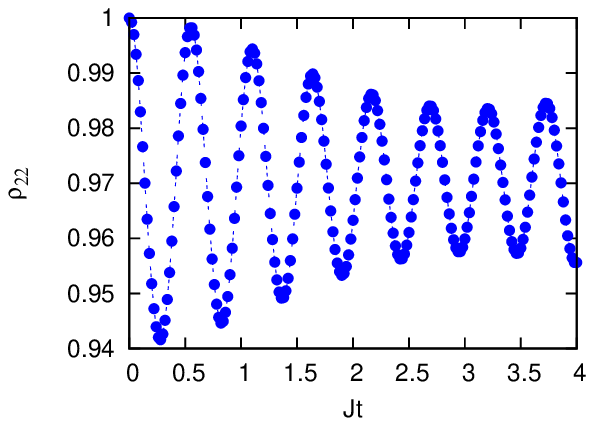}}}
\par}
\caption{The elements of the reduced density matrix corresponding to the initial ground state  for: $h_z=0.5 J$, $h_x=h_y=0$,  $T=2J$. The other parameters are  $S=1/2$, $\omega_0=2J$, $\alpha=\gamma=\lambda=J$. }\label{fig9}
\end{figure}
\begin{figure}[tbh]
{\centering\subfigure[{}]{\resizebox*{0.33\textwidth}{!}{\includegraphics{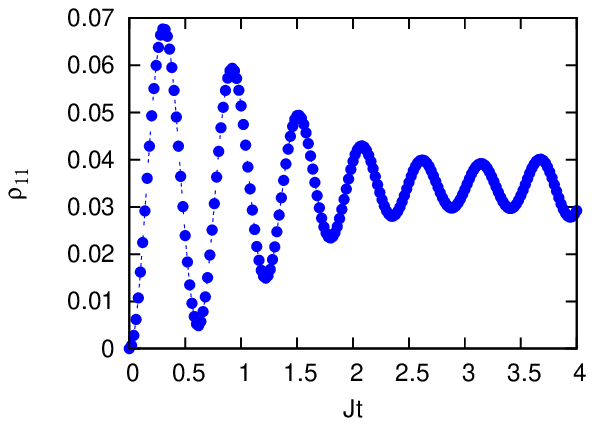}}}
\subfigure[{}]{
\resizebox*{0.33\textwidth}{!}{\includegraphics{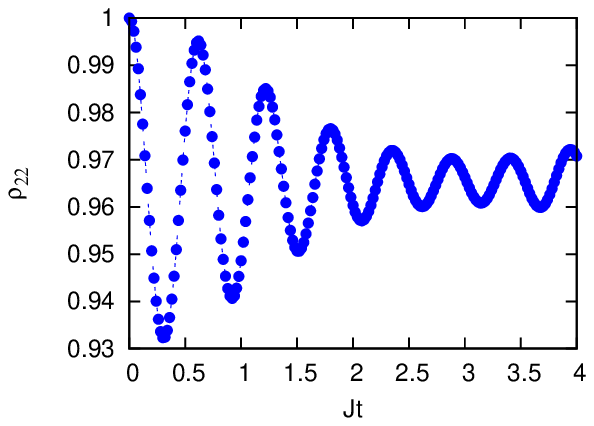}}}
\par}
\caption{The diagonal elements of  the reduced density matrix corresponding to the initial ground state  for: $h_z=0.5 J$, $h_x=h_y=0$,  $T=2.5J$. The other parameters are  $S=1/2$, $\omega_0=2J$, $\alpha=\gamma=\lambda=J$.}\label{fig10}
\end{figure}
\begin{figure}[tbh]
{\centering\subfigure[{}]{\resizebox*{0.33\textwidth}{!}{\includegraphics{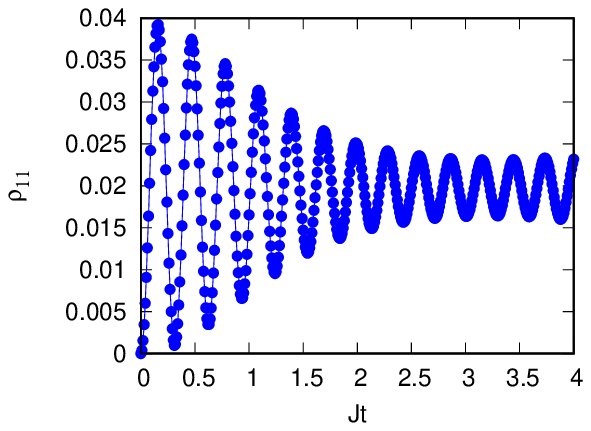}}}
\subfigure[{}]{
\resizebox*{0.33\textwidth}{!}{\includegraphics{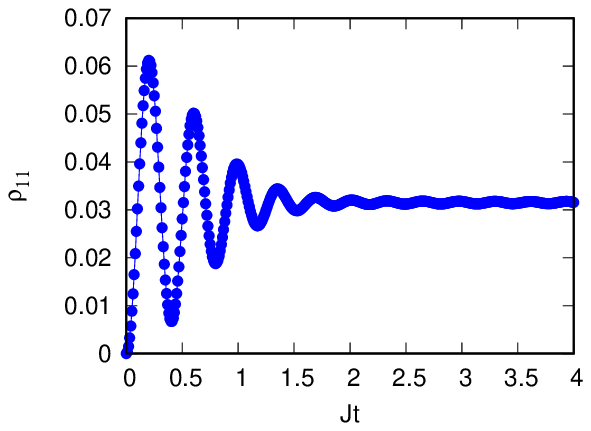}}}
\par}
\caption{The  evolution in time of the element $\rho_{11}$ associated with the initial ground state. For (a): $h_z=0.5 J$, $h_x=h_y=0$,  $T=5J$. For (b): $h_z=0.5 J$, $h_x=h_y=0$,  $T=7J$. The other parameters are  $S=1$, $\omega_0=2J$, $\alpha=\gamma=\lambda=J$.}\label{fig11}
\end{figure}
\begin{figure}[tbh]
{\centering\subfigure[{}]{\resizebox*{0.33\textwidth}{!}{\includegraphics{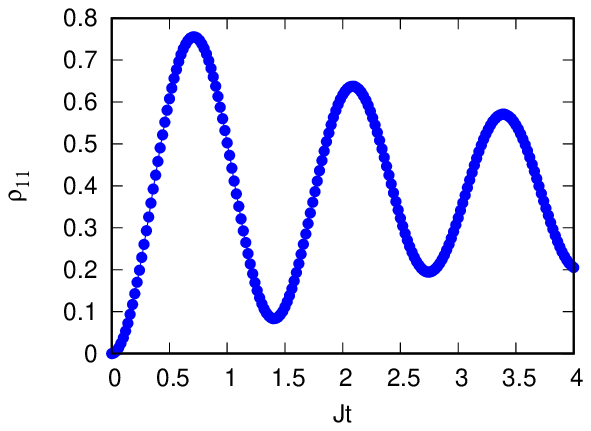}}}
\subfigure[{}]{
\resizebox*{0.33\textwidth}{!}{\includegraphics{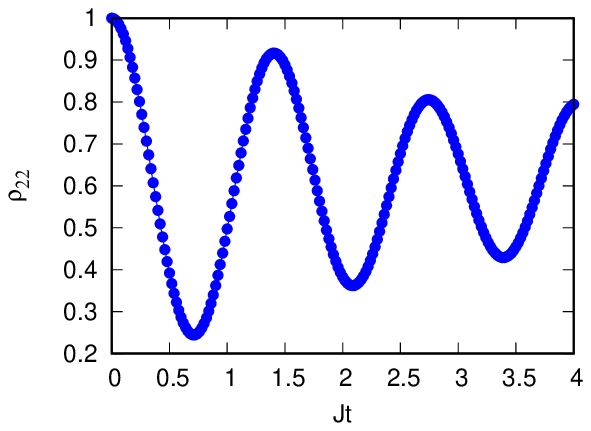}}}
\subfigure[{}]{
\resizebox*{0.33\textwidth}{!}{\includegraphics{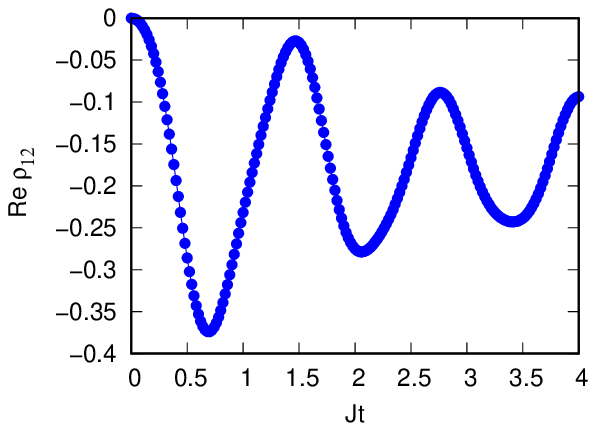}}}
\subfigure[{}]{
\resizebox*{0.33\textwidth}{!}{\includegraphics{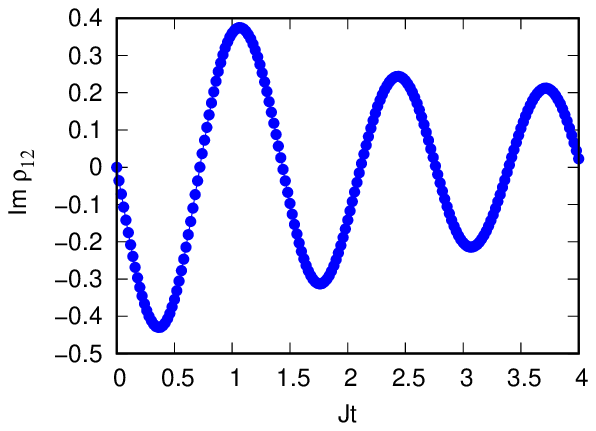}}}
\par}
\caption{The  evolution in time of the element of the reduced density matrix  associated with the initial ground state for:  $h_x=0.5 J$, $h_z=h_y=0$,  $T=2J$. The other parameters are  $S=1/2$, $\omega_0=2J$, $\alpha=\gamma=\lambda=J$.}\label{fig12}
\end{figure}

In order to get more insights into the nature of the dynamics, we introduce, in analogy with the decay rate \cite{petru, ham4},  the so-called dephasing rate
\begin{equation}
 \kappa(t)=-\frac{1}{|F(t)^\eta|}\frac{d}{dt}|F(t)^\eta|.
\end{equation}
From figure \ref{fig8}, we see that the increase of temperature leads to a significant increase of the decay rate when the magnetic field is along the $z$-axis for spin-1/2 but the periods of oscillation remains unchanged. This is to be contrasted to the case where the magnetic field is pointing to the $x$-direction, where we see that the amplitude does not increase much but there occurs a slight decrease of the frequency of oscillation. For $S=1$, when the magnetic field is directed along the $z$-axis, we see that the decay rate is not much affected by the change of the temperature, whereas the frequency of oscillations decreases when the magnetic field points to the $x$-direction. Nevertheless, the non-Markovian character is clearly noticed from the oscillation of the decay rates between positive and negative values. This manifestation is more noticeable when the magnetic field has no component along the $z$ axis.  
\section{ Energy level transition \label{Sec4}}
Consider now the  case  where the atom is initially prepared in its ground state
\begin{equation}
 \phi(0)=|g\rangle,
\end{equation}
and is isotropically coupled to the atoms of the lattice, that is, we take $\alpha=\delta=\gamma$.
We shall  investigate the effect of the relevant parameters of the model on the exited state population, i.e., we shall study the probability of transition of the atom  to the excited state. The latter is best described by the matrix element $\rho_{11}$ or equivalently by the component $r_3$. The transition occurs as a result of  the interaction of the atom with the spins of the lattice, leading to energy exchange between the two systems, which affects both the diagonal and the off-diagonal elements of the reduced density matrix.

When the magnetic field points along the z-direction, the excited state population is given by

\begin{eqnarray}
\rho_{11}(t)=\frac{1}{\tilde Z} \sum\limits_{j=0,1/2}^{\eta S} \alpha^2 \nu(j,\eta; S) \sum\limits_{\ell=-j}^{j} (j(j+1)-\ell(\ell-1))e^{\beta (12 J m+h_z)\ell}\nonumber \\
\times \sin^2\bigl[\tfrac{t}{2} \sqrt{  M_-(j,\ell)}\bigr]/  \sqrt{{  M_-(j,\ell)}}
\end{eqnarray}
where
\begin{equation}
  M_-(j,\ell)=\alpha^2(j(j+1)-\ell(\ell-1))+\frac{1}{4}\bigl[2(h_z+12 J m+\omega_0)+\alpha(2\ell-1)\bigr]^2,
\end{equation}
and
\begin{equation}
\tilde Z= \sum\limits_{j=0}^{\eta S}  \nu(j,\eta; S)\sum\limits_{\ell=-j}^{j} e^{\beta (12 J m+h_z)\ell}.
\end{equation}
The off-diagonal element remains always zero, i.e. $\rho_{12}(t)=0$. Notice that in the above equation $ \nu(j,\eta; S)$ denotes the degeneracy corresponding to the angular momentum $j$ resulting from the addition of the spin operators of the unit cell. It is given by \cite{ham5}:
\begin{eqnarray}
 \nu(j,\eta;S)=\sum\limits_{L_{-S+2},L_{-S+3},\cdots, L_{S}=0}^\eta\frac{\eta!}{\prod\limits_{k=2}^{2S} (L_{-S+k})!} \Biggl(\frac{(2S-1)\eta+2j+1-
 \sum\limits_{\rho=2}^{2S}(2\rho-1)L_{-s+\rho}}
 {S\eta+j+1-\sum\limits_{\rho=2}^{2S}\rho \ L_{-s+\rho}}\Biggr)\nonumber \\
 \times \Biggl[\Bigl(S\eta+j-\sum\limits_{\rho=2}^{2S}\rho \ L_{-s+\rho}\Bigr)!\Bigl((1-S)\eta-j+\sum\limits_{\rho=2}^{2S}(\rho-1) L_{-s+\rho}\Bigr)!\Biggr]^{-1}\label{deg}
\end{eqnarray}
where the integers $L_{i}$ assume values ranging from 0 to $\eta$.
\begin{figure}[tbh]
{\centering\subfigure[{}]{\resizebox*{0.33\textwidth}{!}{\includegraphics{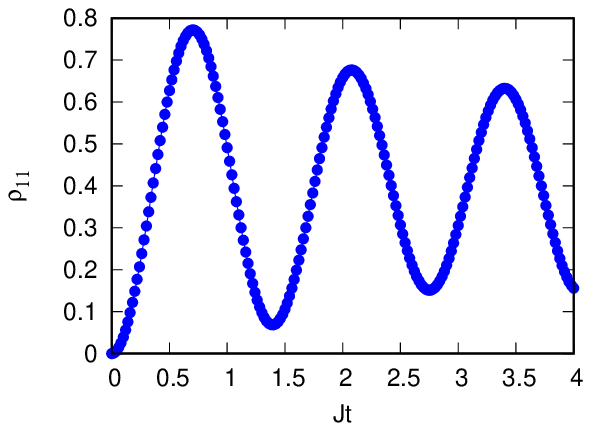}}}
\subfigure[{}]{
\resizebox*{0.33\textwidth}{!}{\includegraphics{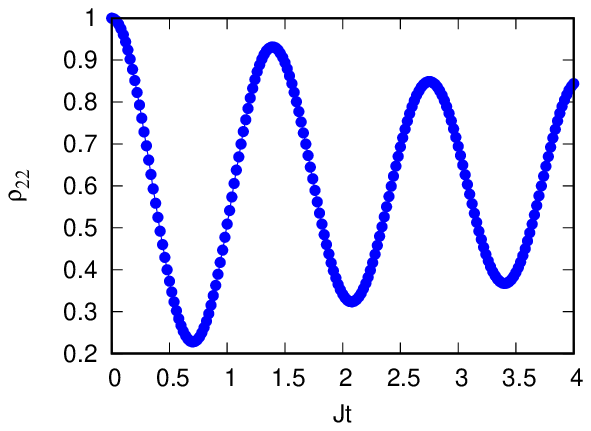}}}
\subfigure[{}]{
\resizebox*{0.33\textwidth}{!}{\includegraphics{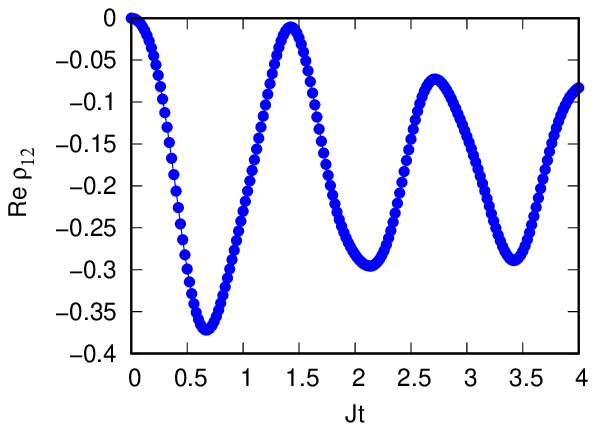}}}
\subfigure[{}]{
\resizebox*{0.33\textwidth}{!}{\includegraphics{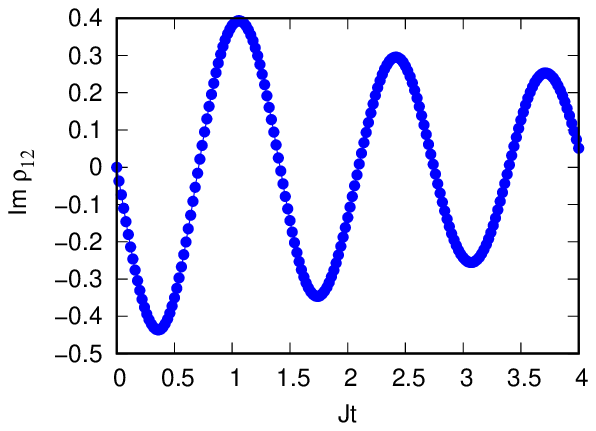}}}
\par}
\caption{The same as figure \ref{fig12} but with $h_x=J$ and $T=2J$.}\label{fig13}
\end{figure}
\begin{figure}[tbh]
{\centering\subfigure[{}]{\resizebox*{0.33\textwidth}{!}{\includegraphics{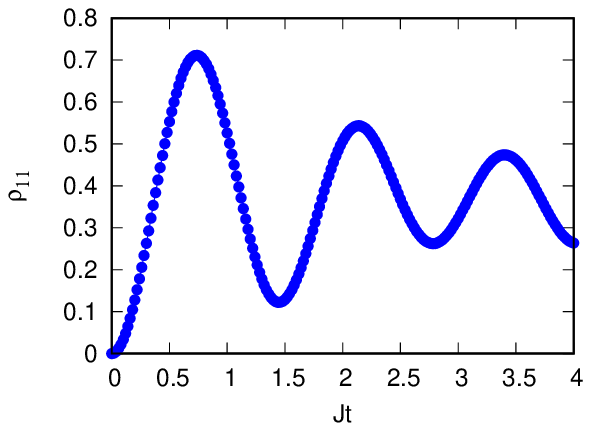}}}
\subfigure[{}]{
\resizebox*{0.33\textwidth}{!}{\includegraphics{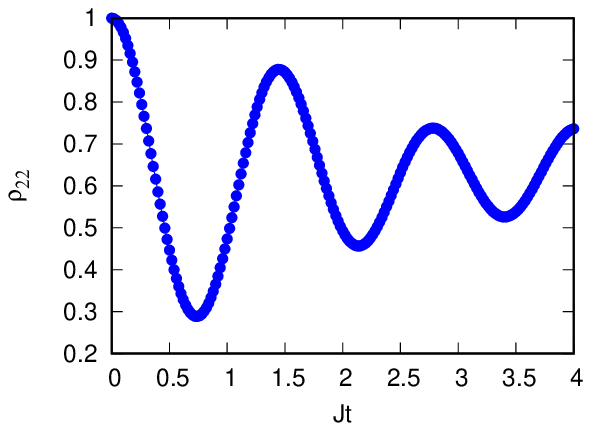}}}
\subfigure[{}]{
\resizebox*{0.33\textwidth}{!}{\includegraphics{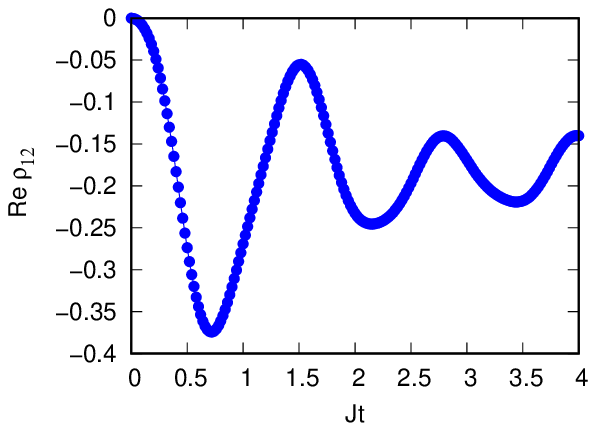}}}
\subfigure[{}]{
\resizebox*{0.33\textwidth}{!}{\includegraphics{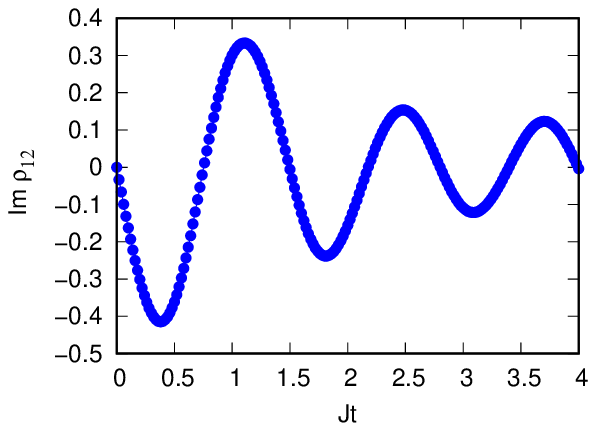}}}
\par}
\caption{The same as figure \ref{fig13} but with  $T=2.5J$.} \label{fig14}
\end{figure}

In figures \ref{fig9} and \ref{fig10}, we display the variation in time of the elements of the reduced density matrix when the magnetic field is directed along the $z$-axis. We notice that there exists a non-zero, yet small,  probability that the atom occupies its excited state, which slightly increases  with the temperature.  It turns out that changing the strength of the magnetic field does not affect much the transition probability. On the other hand, the state of the atom remains mixed at all moments. The same occurs  when the lattice is composed of spin-1 particles, but in this case we observe a slight decrease of the transition probability as depicted in Fig.~\ref{fig11} in contrast to one might expect at higher temperatures. 

The probability of transition significantly increases when the magnetic field points in the $x$-direction, as illustrated in Figs. \ref{fig12}-\ref{fig14} for a lattice of spin-1/2. In this case the periods of oscillations  are  longer as compared with those in figures \ref{fig9} and \ref{fig10}. Increasing the strength of the magnetic field induces slightly larger probability occupation of the excited state. This behavior is also produced when the magnetic field points in the $y$ direction. However, in contrast to the case where the magnetic field is parallel to the $z$-axis, we see that here, as the temperature increases, the occupation probability decreases.  The real part of $\rho_{12}$ remains negative, while its imaginary part oscillates about zero. The opposite is true when the magnetic field points in the $y$-direction. Generally speaking, it is  found that the more the magnetic field deviates from the $x$-$y$ plane the smaller the occupation probability is.

\begin{figure}[tbh]
{\centering\subfigure[{}]{\resizebox*{0.33\textwidth}{!}{\includegraphics{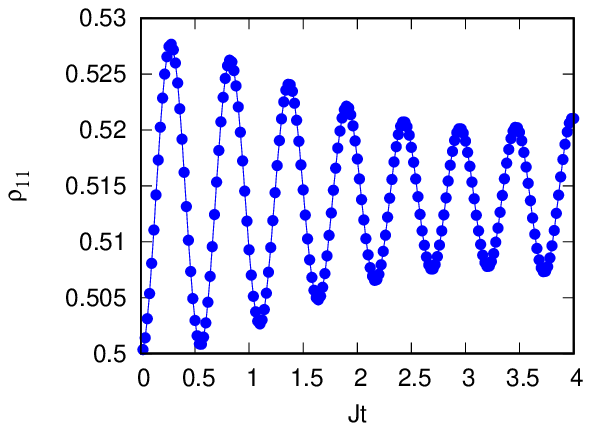}}}
\subfigure[{}]{
\resizebox*{0.33\textwidth}{!}{\includegraphics{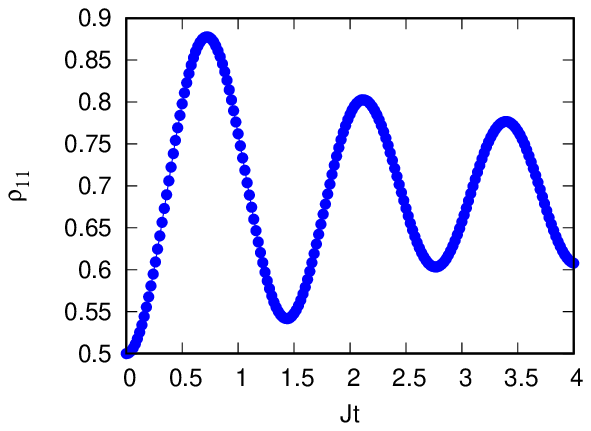}}}
\par}
\caption{The excited state population as a function of time corresponding to the initial state $\phi_s(0)$ with: (a) $h_z=0.5 J$, $h_x=h_y=0$, and (b) $h_x=0.5 J$, $h_x=h_y=0$. The other parameters are $T=2 J$, $S=1/2$, $\omega_0=2J$, $\alpha=\gamma=\lambda=J$.} \label{fig15}
\end{figure}

\begin{figure}[tbh]
{\centering\subfigure[{}]{\resizebox*{0.33\textwidth}{!}{\includegraphics{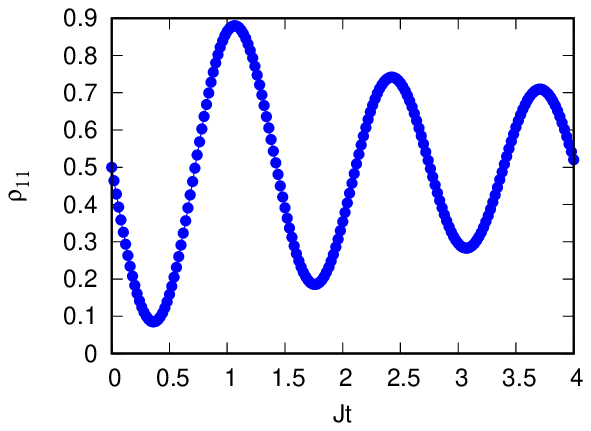}}}
\subfigure[{}]{
\resizebox*{0.33\textwidth}{!}{\includegraphics{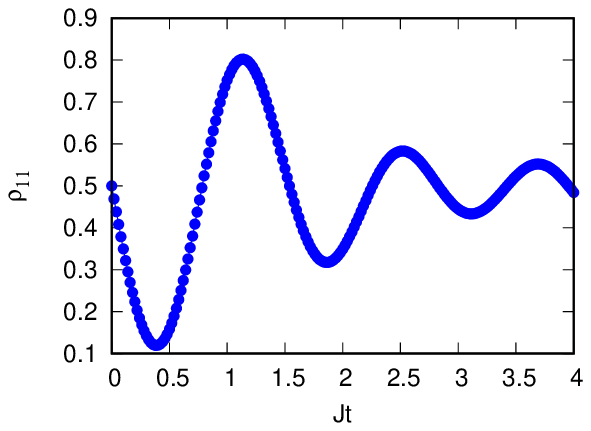}}}
\par}
\caption{The excited state population as a function of time corresponding to the initial state $\phi_s(0)$ with: (a) $h_y=0.5 J$,  $T=2J$, $h_x=h_z=0$, and (b) $h_y=0.5 J$, $h_x=h_z=0$, $T=2.5 J$. The other parameters are $S=1/2$, $\omega_0=2J$, $\alpha=\gamma=\lambda=J$.} \label{fig16}
\end{figure}

Let us consider now the instance in which the initial state of the atom is the state:
\begin{equation}
  \phi_s(0)=\frac{1}{\sqrt{2}}(|g\rangle+|e\rangle).
 \end{equation}
 The initial occupation probability is the same for both the exited and the ground state. We find that the atom is most likely to be found in the excited state (see figure \ref{fig15}) unless the magnetic field is pointing in the $y$ direction where the excited state occupation probability may decrease compared to its initial value as illustrated in Fig. \ref{fig16}. Hence the atom periodically undergoes transitions between these states, which persist at longer times due to the non-Markovian nature of the dynamics.

To assess the effect of the spin magnitude when the magnetic field points in the $x$-direction, we suppose that the states $|g\rangle$ and $|e\rangle$ are degenerate, which amounts to choosing $\omega_0=0$. When the central system is initially prepared in the state $|g\rangle$, then the evolution in time of the reduced density matrix element $\rho_{11}$ is given by 
\begin{eqnarray}
 \rho_{11}(t)=\frac{1}{2}\bigl[ 1- {\rm Re} \Psi(t)\bigr],
\end{eqnarray}
with
\begin{eqnarray}
   \Psi(t)&=& \frac{1}{\tilde Z} \sum\limits_{j=0}^{\eta S} \nu(j,\eta; S) \sum\limits_{\ell=-j}^{j} e^{(it+\beta \ell)(12Jm+h_x)}\nonumber \\ &\times&  \Bigl[\cos\bigl[\tfrac{t}{2} \sqrt{  M_+(j,\ell)}\bigr]-i\frac{(12Jm+h_x)+\alpha (\ell+\frac{1}{2})}{\sqrt{  M_+(j,\ell)}} \sin\bigl[\tfrac{t}{2} \sqrt{  M_+(j,\ell)}\bigr]   \Bigr]
    \\ &\times&  \Bigl[\cos\bigl[\tfrac{t}{2} \sqrt{  M_-(j,\ell)}\bigr]-i\frac{(12J m+h_x)+\alpha (\ell-\frac{1}{2})}{\sqrt{  M_-(j,\ell)}} \sin\bigl[\tfrac{t}{2} \sqrt{  M_-(j,\ell)}\bigr]   \Bigr]
\end{eqnarray}
where $   M_-(j,\ell)$ is defined as above, and
\begin{equation}
  M_+(j,\ell)=\alpha^2(j(j+1)-\ell(\ell+1))+\frac{1}{4}\bigl[2(h_x+12 J m+\omega_0)+\alpha(2\ell+1)\bigr]^2.
  \end{equation}
  \begin{figure}[tbh]
{\centering\subfigure[{}]{\resizebox*{0.33\textwidth}{!}{\includegraphics{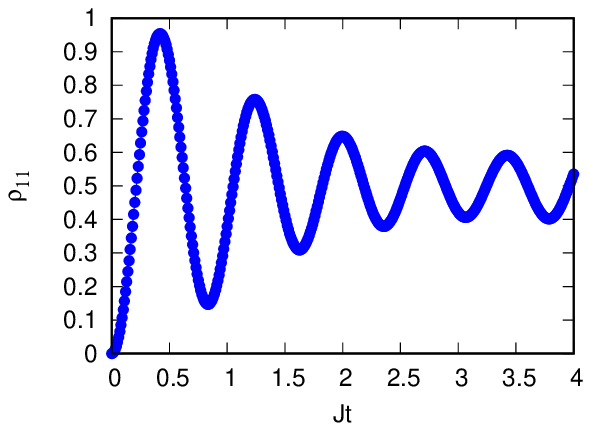}}}
\subfigure[{}]{
\resizebox*{0.33\textwidth}{!}{\includegraphics{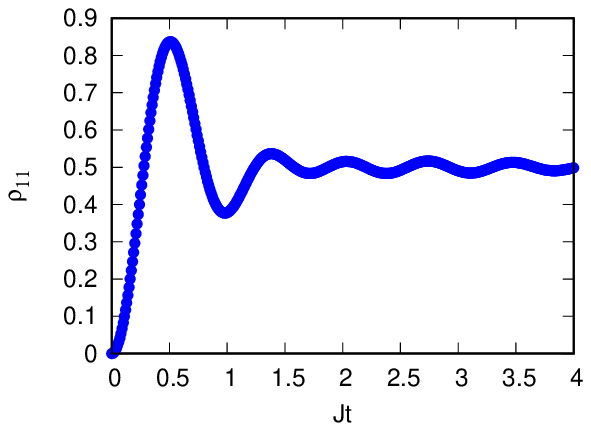}}}
\par}
\caption{The excited state population as a function of time corresponding to the initial ground  state $\phi(0)$ with: (a) $h_x=0.5 J$,  $T=5J$, $h_y=h_z=0$, and (b) $h_x=0.5 J$, $h_y=h_z=0$, $T=7J$. The other parameters are $S=1$, $\omega_0=0$, $\alpha=\gamma=\lambda=J$.} \label{fig17}
\end{figure}
It is clearly seen from figure \ref{fig17} that the system evolves at high temperatures  to a state where the ground and the excited state have equal statistical weights. This occurs in  much faster rate  as compared to the evolution in a lattice of spin-1/2 particles.
\section{Entanglement of two atoms located at non-adjacent cells \label{sec5}}
Let us consider two  two-level atoms occupying the centers of non-adjacent unit cells in the lattice,  such that each cell does not share any  spin with the other one. In such situation, the effect of the lattice on either atom is reduced to the coupling to its own  local unit cell. We further assume that  the two-level systems  do not interact with each other and  that  they are  initially prepared in the Bell maximally entangled state:

\begin{equation}
  |\Phi\rangle=\frac{1}{\sqrt{2}}(|gg\rangle+|ee\rangle).
 \end{equation}
The initial density matrix corresponding to the latter state  may be written as 
\begin{eqnarray}
 \rho(0)\equiv|\Phi\rangle\langle \Phi|=\frac{1}{4}\mathbb{I}_4+\frac{1}{4}\sigma_z\otimes \sigma_z+\frac{1}{2}\bigl[\sigma_+\otimes \sigma_++\sigma_-\otimes \sigma_-\Bigr].
\end{eqnarray}

When the magnetic field points in the $z$-direction, we find after tracing out the lattice degrees of freedom that the density matrix evolves according to
\begin{equation}
\rho(t)= \begin{pmatrix}
  \frac{1}{4}+\frac{\Xi_+(t)^2}{4} && 0 && 0 && \frac{\Psi(t)^2}{2}\\
  0 &&  \frac{1}{4}+\frac{\Xi_+(t) \Xi_-(t)}{4} && 0 &&0\\
  0 && 0 &&  \frac{1}{4}+\frac{\Xi_+(t) \Xi_-(t)}{4} && 0 \\
   \frac{{\Psi^*(t)}^2}{2} && 0 && 0&&  \frac{1}{4}+\frac{\Xi_-(t)^2}{4}
 \end{pmatrix},
\end{equation}
where
\begin{eqnarray}
 \Xi_\pm(t)&= &\mp 1 \pm \frac{2}{\tilde Z} \sum\limits_{j=0}^{\eta S} \nu(j,\eta; S) \sum\limits_{\ell=-j}^{j} e^{\beta \ell(12Jm+h_z)}\nonumber \\ &\times&  \Bigl\{\cos\bigl[\tfrac{t}{2} \sqrt{  M_\pm(j,\ell)}\bigr]^2-\frac{\Bigl((12J m +h_z)+\alpha (\ell\pm\frac{1}{2})\Bigr)^2}{  M_\pm(j,\ell)} \sin\bigl[\tfrac{t}{2} \sqrt{  M_\pm(j,\ell)}\bigr]^2   \Bigr\}   .  
\end{eqnarray}
The amount of  entanglement of a state $\rho$ (whether mixed or not) of a bipartite  two-level system   may be quantified  using the concurrence $C(\rho)$ \cite{wootters,vidal}, which is defined by
\begin{equation}
 C(\rho)={\rm max}\{0, \sqrt{\nu_1}- \sqrt{\nu_2}- \sqrt{\nu_3}- \sqrt{\nu_4}\},
\end{equation}
where $ \nu_1$, $\nu_2$, $ \nu_3$ and $\nu_4$ are the eigenvalues, in descending order, of the operator
\begin{equation}
\tilde\rho(t) = \rho(t)(\sigma_y \otimes \sigma_y )\rho^*(t)(\sigma_y \otimes \sigma_y ).
\end{equation}
The concurrence takes on values ranging from zero for separable states to one for maximally entangled states. 
\begin{figure}[tbh]
{\centering\subfigure[{}]{\resizebox*{0.33\textwidth}{!}{\includegraphics{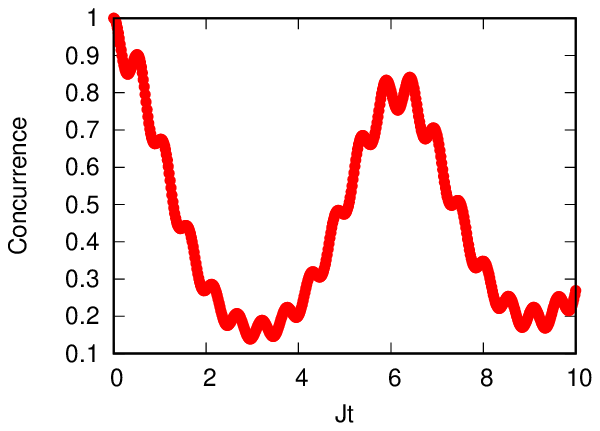}}}
\subfigure[{}]{
\resizebox*{0.335\textwidth}{!}{\includegraphics{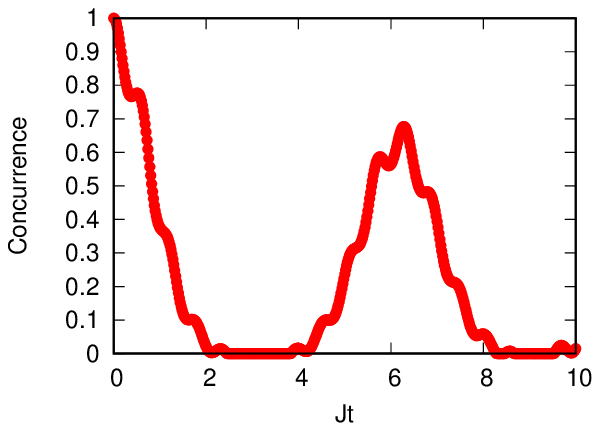}}}
\par}
\caption{The concurrence  as a function of time corresponding to the initial state $|\Phi\rangle$ with: (a) $h_z=0.5 J$,  $T=2J$, $h_x=h_y=0$, and (b) $h_z=0.5 J$, $h_x=h_y=0$, $T=2.5 J$. The other parameters are $S=1/2$, $\omega_0=2J$, $\alpha=\gamma=\lambda=J$.} \label{fig18}
\end{figure}

\begin{figure}[tbh]
{\centering\subfigure[{}]{\resizebox*{0.33\textwidth}{!}{\includegraphics{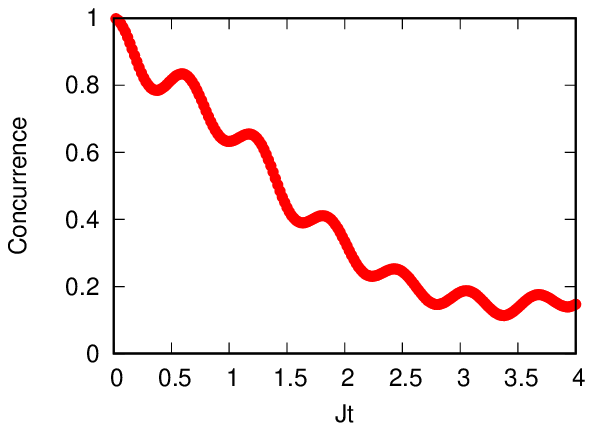}}}
\subfigure[{}]{
\resizebox*{0.335\textwidth}{!}{\includegraphics{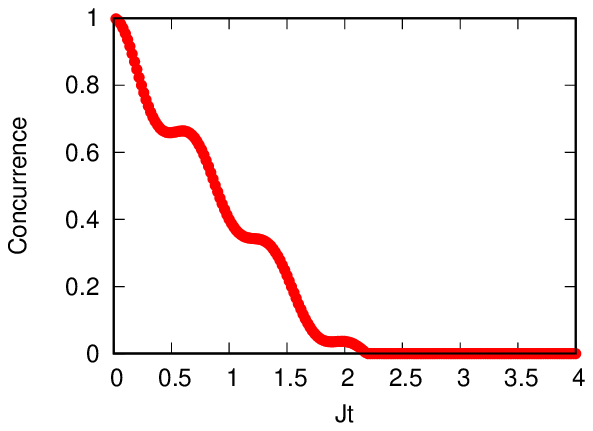}}}
\par}
\caption{The concurrence  as a function of time corresponding to the initial state $|\Phi\rangle$ with: (a) $h_x=0.5 J$,  $T=2J$, $h_y=h_z=0$, and (b) $h_x=0.5 J$, $h_y=h_z=0$, $T=2.5 J$. The other parameters are $S=1/2$, $\omega_0=2J$, $\alpha=\gamma=\lambda=J$.} \label{fig19}
\end{figure}
 
\begin{figure}[tbh]
{\centering\subfigure[{}]{\resizebox*{0.33\textwidth}{!}{\includegraphics{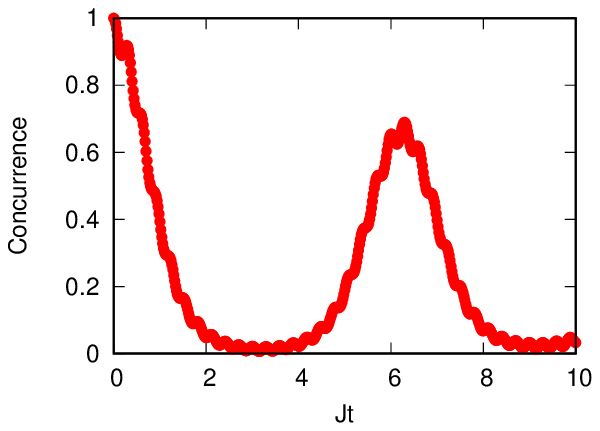}}}
\subfigure[{}]{
\resizebox*{0.335\textwidth}{!}{\includegraphics{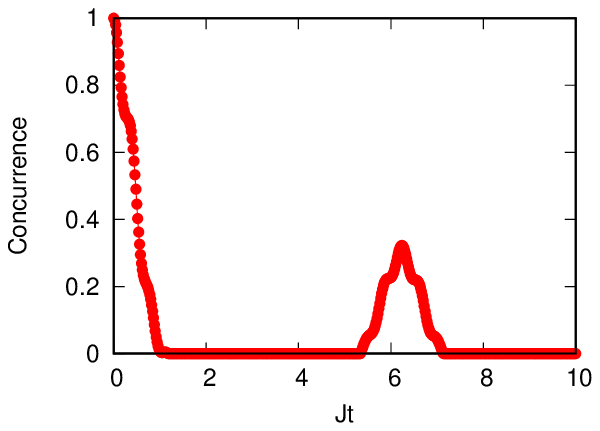}}}
\par}
\caption{The concurrence  as a function of time corresponding to the initial state $|\Phi\rangle$ with: (a) $h_z=0.5 J$,  $T=5J$, $h_x=h_z=0$, and (b) $h_z=0.5 J$, $h_x=h_z=0$, $T=7J$. The other parameters are $S=1$, $\omega_0=2J$, $\alpha=\gamma=\lambda=J$.} \label{fig20}
\end{figure}

\begin{figure}[tbh]
{\centering\subfigure[{}]{\resizebox*{0.4\textwidth}{!}{\includegraphics{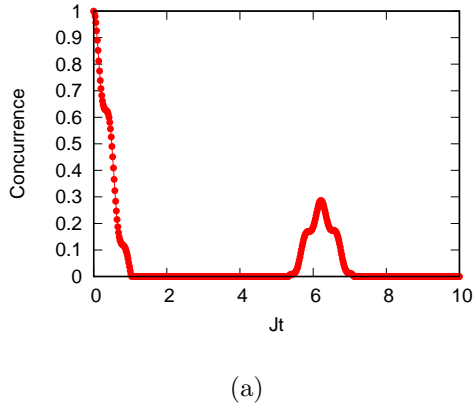}}}
\par}
\caption{The concurrence  as a function of time corresponding to the initial state $|\Phi\rangle$ for $h_x=0.5 J$,  $T=7J$, $h_y=h_z=0$,  $S=1$, $\omega_0=0$, $\alpha=\gamma=\lambda=J$.} \label{fig21}
\end{figure}

In figures \ref{fig18}-\ref{fig19} we display the variation in time of the concurrence for different values of the magnetic field and  the temperature for a lattice of spin-1/2. It can be seen that close to the critical temperature there occurs a sudden death of the entanglement for both directions of the magnetic field. The decay and revival of the entanglement occur at comparable rates. Nevertheless, we can see that while the evolution of the entanglement   is less  dependent on the direction of the magnetic field it is very sensitive to the temperature, in contrast to the dephasing and the population evolution as discussed above. Again, the complete loss of entanglement and its revival is mainly due to the competition between thermal fluctuations and the memory effects of the lattice responsible for the back flow of information from the latter to the central atoms. For a lattice of spin-1 particles, we see from figure \ref{fig20} that the decay of entanglement becomes much faster and its revival takes much longer times when the magnetic field points in the $z$-direction.       
In case where  $\omega_0=0$, and the magnetic field points in the $x$-direction, the reduced density matrix is calculated as:
\begin{equation}
\rho(t)= \begin{pmatrix}
  \tfrac{1+{\rm Re} \Psi(t)^2}{4} && \frac{\Delta(t)-4i {\rm Im } \Psi(t)^2}{16} &&  \frac{\Delta(t)-4i {\rm Im } \Psi(t)^2}{16} &&  \tfrac{\Xi_+(t)^2\Xi_-(t)^2+{\rm Re} \Psi(t)^2}{4}\\
   \frac{\Delta(t)+4i {\rm Im } \Psi(t)^2}{16} && \tfrac{1-{\rm Re} \Psi(t)^2}{4} && \tfrac{\Xi_+(t)^2\Xi_-(t)^2-{\rm Re} \Psi(t)^2}{4} &&  \frac{\Delta(t)+4i {\rm Im } \Psi(t)^2}{16}\\
   \frac{\Delta(t)+4i {\rm Im } \Psi(t)^2}{16} &&  \tfrac{\Xi_+(t)^2\Xi_-(t)^2-{\rm Re} \Psi(t)^2}{4} && \tfrac{1-{\rm Re} \Psi(t)^2}{4}  &&   \frac{\Delta(t)+4i {\rm Im } \Psi(t)^2}{16} \\
   \tfrac{\Xi_+(t)^2\Xi_-(t)^2+{\rm Re} \Psi(t)^2}{4} &&  \frac{\Delta(t)-4i {\rm Im } \Psi(t)^2}{16} &&  \frac{\Delta(t)-4i {\rm Im } \Psi(t)^2}{16}&& \tfrac{1+{\rm Re} \Psi(t)^2}{4}
   \end{pmatrix},
\end{equation}
where $ \Delta(t)=\Xi_+(t)^2-\Xi_-(t)^2.$ It turns out (see Fig.~\ref{fig21}) that the entanglement displays here very similar behavior as  that corresponding to a magnetic field that is directed along the $z$-direction, which confirms once again the results obtained above for spin-1/2.

\section{Conclusion} In conclusion we employed the meant-field approximation to study the dynamics of a two level atom that is coupled to the spins in a ferromagnetic lattice. This approximation along with the translation invariance of the lattice enabled us to reduce the $N+1$ body problem to that of $\eta+1$  problem where $\eta$ is the number of spins in a unit cell. It is found that  depending on the orientation of the applied magnetic field various features of the dynamics may occur. Specifically, for the dephasing dynamics, when the magnetic field points in the $z$ direction, the revival of the coherences is less important compared to when the field points in the $x$ or $y$ directions. In the former case the coherence are very sensitive to the variation of the temperature, whereas in the latter case,  they are quite robust with respect to thermal fluctuations. When the lattice spins are equal to one, we found a clear increase of the rate of the loss of the coherence when the magnetic field is along the $z$-axis as compared to the spin one-half case. When the atom is prepared in its ground state, a magnetic field along the $z$ direction  produces a small probability of transition to the excited state. This transition is rather enhanced by the thermal fluctuations. On the contrary, these fluctuations  reduce the occupation probability of the excited state which turns out to be quite important when the magnetic field lies in the $x$-$y$ plane. The concurrence is used to quantify the entanglement  of two atoms located at non-adjacent cells that are initially prepared in the Bell maximally entangled state. Its evolution is rather the same for all directions of the magnetic field. However  it displays entanglement sudden death and revival close to the Currie temperature in either direction.  All the obtained results show clearly the importance of memory effects for temperatures bellow the Curie temperature where the dynamics display non-Markovian features.


\begin{thebibliography}{99}
\bibitem{kittel} C.~Kittel, {\it Introduction to Solid State Physics} (Wiley, New York, 1953).
\bibitem{loss}
D.~P.~DiVincenzo and D.~Loss, J. Magn. Magn. Matter. \textbf{200}, 202 (1999).
\bibitem{loss2}
D.~ Loss and D.~P.~ DiVincenzo Phys.\ Rev.\ A \textbf{57}, 120(1998).
\bibitem{loss3}
G.~Burkard, D.~Loss, and D.~P.~DiVincenzo , Phys.\ Rev.\ B \textbf{59}, 2070 (1999).
\bibitem{nielsen}
M.~A.~Nielsen and I.~L.~Chuang,
\emph{Quantum Computation and Quantum Information}
(Cambridge University Press, Cambridge, 2000).
\bibitem{shor}
P.~W.~ Shor, Phys.\ Rev.\ A \textbf{52}, R2493 (1995).
\bibitem{ekert}
A. Barenco, D. Deutsch, A. Ekert, and R. Jozsa, Phys. Rev. Lett. \textbf{74}, 4083 (1995).
\bibitem{cirac}
J. I. Cirac and P. Zoller, Phys. Rev. Lett. \textbf{74}, 4091 (1995).
\bibitem{mosca}
J. A. Jones, M. Mosca, and R. H. Hansen, Nature \textbf{393}, 344 (1995).
\bibitem{chuang}
I. L. Chuang, N. Gershenfeld, and M. Kubinec, Phys. Rev. Lett \textbf{80}, 3408 (1998).
\bibitem{zurek1}
W.~H.~Zurek, Phys.\ Today \textbf{44}, No. 10, 36 (1991).
\bibitem{petru}
H.~P.~Breuer and F.~Petruccione,
\emph{The Theory of Open  Quantum Systems}
(Oxford University Press, Oxford, 2002).

\bibitem{kah}
A. V. Khaetskii, D. Loss, and L. Glazman, Phys. Rev.Lett. \textbf {88}, 186802 (2002).
\bibitem{burg}
H.~P.~Breuer, D.~Burgarth and F.~Petruccione,
Phys.\ Rev.\ B \textbf{70}, 045323 (2004).
\bibitem{bose}
A. Hutton and S. Bose, Phys. Rev. A \textbf{69}, 042312 (2004).
\bibitem{paga}
M. Lucamarini, S. Paganelli, and S. Mancini, Phys. Rev. A \textbf{69}, 062308 (2004).
\bibitem{zurek2}
F. M. Cucchietti, J. P. Paz, and W. H. Zurek, Phys. Rev. A \textbf{72}, 052113 (2005)
\bibitem{ham1}
Y.~Hamdouni, M.~Fannes, and F.~Petruccione, Pys.\ Rev.\ B \textbf{73}, 245323 (2006).
\bibitem{quan}
H. T. Quan, Z. Song, X. F. Liu, P. Zanardi, and C. P. Sun, Phys. Rev. Lett. \textbf{96}, 140604 (2006).
\bibitem{yuan}
X. Z. Yuan, H. S. Goan, and K. D. Zhu, Phys. Rev. B \textbf{75}, 045331 (2007).
\bibitem{fischer}
J. Fischer and H.-P. Breuer, Phys. Rev. A \textbf {76}, 052119 (2007).
\bibitem{ham2}
Y. Hamdouni and F. Petruccione, Phys. Rev. B \textbf{76}, 174306 (2007).
\bibitem{petru2}
H.-P. Breuer and F. Petruccione, Phys. Rev. E \textbf {76}, 016701 (2007).
\bibitem{lai}
C. Y. Lai, J. T. Hung, C. Y. Mou, and P. Chen, Phys. Rev. \textbf{B} 77, 205419 (2008).
\bibitem{taka}
S. Takahashi, R. Hanson, J. van Tol, M. S. Sherwin, and D. D. Awschalom, Phys. Rev. Lett. \textbf{101}, 047601 (2008).
\bibitem{sarma}
W. M. Witzel and S. Das Sarma, Phys. Rev. B \textbf{ 77}, 165319 (2008).
\bibitem{lu}
P. Lu, H.-L. Shi, L. Cao, X.-H. Wang, T. Yang, J. Cao, and W.-L. Yang, Phys. Rev. B \textbf {101}, 184307 (2020).
\bibitem{zeji}
Z. Li, P. Yang, W.-L. You, and N. Wu, Phys. Rev. A \textbf{102} 032409 (2020).
\bibitem{ham3}
Y. Hamdouni, J. Phys. A: Math. Theo. \textbf{40}, 11569 (2007); Y. Hamdouni, J. Phys. A: Math. Theo. \textbf{42}, 315301 (2009);
Y. Hamdouni, Phys. Lett. A \textbf{373}, 1233 (2009); Y. Hamdouni, J. Phys. A: Math. Theo. \textbf{45}, 425301 (2012).
\bibitem{kici}
M. Kici\'nski and J. K. Korbicz, Phys. Rev. A \textbf{104}, 042216 (2021).
\bibitem{li}
L. Jiaxiu, C. Ye, and W. Ning, Phys. Rev. B \textbf{106}, 024302 (2022).

\bibitem{rivas}
A. Rivas, S. F. Huelga, and M. B. Plenio, Phys. Rev. Lett. \textbf{105}, 050403 (2010).
\bibitem{lu2}
X.-M. Lu, X. Wang, and C. P. Sun, Phys. Rev. A \textbf{82}, 042103 (2010).
\bibitem{laine}
E.-M. Laine, J. Piilo, and H.-P. Breuer, Phys. Rev. A \textbf{81}, 062115 (2010).
\bibitem{vasile}
R. Vasile, S. Maniscalco, M. G. A. Paris, H.-P. Breuer, and J. Piilo, Phys. Rev. A \textbf{84}, 052118 (2011).
\bibitem{fu}
S. Luo, S. Fu, and H. Song, Phys. Rev. A \textbf{86}, 044101 (2012).
\bibitem{hall}
M. J. W. Hall, J. D. Cresser, L. Li, and E. Andersson, Phys. Rev. A \textbf{89}, 042120 (2014).
 \bibitem{mahan}
 G. D. Mahan, {\it Many-Particle Physics}, 2nd ed (Plenum Press, New York, 1990).
\bibitem{majlis}
N.~Majlis, {\it The Quantum Theory of Magnetism} (World Scientific, New Jersey, 2007).
\bibitem{ham4} Y. Hamdouni,  Phys. Rev. A \textbf{103}, 012209 (2021).
\bibitem{ham5} Y. Hamdouni, Phys. Rev. A \textbf{94}, 022120 (2016).
\bibitem{wootters} 
W. K. Wootters, Phys. Rev. Lett. \textbf{80}, 2245 (1998).
\bibitem{vidal}
G. Vidal, Phys. Rev. A \textbf{62}, 062315 (2000).

 

\end{thebibliography}
\end{document}